\documentclass{article} 
\usepackage{iclr2025_conference,times}

\usepackage{multirow}
\usepackage{graphicx}
\usepackage{subcaption}

\usepackage[T1]{fontenc}    
\usepackage{hyperref}       
\usepackage{url}            
\usepackage{booktabs}       
\usepackage{amsfonts}       
\usepackage{nicefrac}       
\usepackage{microtype}      
\usepackage{amsmath}
\usepackage{algorithm} 
\usepackage{wrapfig}
\usepackage{algpseudocode} 
\usepackage{bbm}
\usepackage{pifont}
\usepackage{graphicx}
\usepackage{amssymb}
\usepackage{caption}
\usepackage{subcaption}
\usepackage{multirow}
\usepackage{colortbl}
\usepackage[most]{tcolorbox}
\newtcolorbox{response}[1][]{
  colback=gray!5,
  colframe=black,
  fonttitle=\bfseries,
  coltitle=black,
  }
\definecolor{redshade}{HTML}{e6e6e6}
\definecolor{greyshade}{HTML}{FF0000}

\title{Self-Evolving Multi-Agent Collaboration Networks for Software Development}

\author{Yue Hu\textsuperscript{1}, Yuzhu Cai\textsuperscript{2,3}, Yaxin Du\textsuperscript{1}, Xinyu Zhu\textsuperscript{1}, 
\textbf{Xiangrui Liu}\textsuperscript{1}, \textbf{Zijie Yu}\textsuperscript{1}, \\  \textbf{Yuchen Hou}\textsuperscript{1},  \textbf{Shuo Tang}\textsuperscript{1},  \textbf{Siheng Chen}\textsuperscript{1,3} 
\\
\textsuperscript{1} Shanghai Jiao Tong University, \textsuperscript{2} Beihang University, \textsuperscript{3} Shanghai AI Laboratory
\\
\textsuperscript{1} \texttt{\{1867112931,sihengc\}@sjtu.edu.cn,} \textsuperscript{2} \texttt{caiyuzhu@buaa.edu.cn} 
}

\iclrfinalcopy 
\begin{document}

\maketitle

\begin{abstract}
LLM-driven multi-agent collaboration (MAC) systems have demonstrated impressive capabilities in automatic software development at the function level. However, their heavy reliance on human design limits their adaptability to the diverse demands of real-world software development.
To address this limitation, we introduce EvoMAC, a novel self-evolving paradigm for MAC networks. Inspired by traditional neural network training, EvoMAC obtains text-based environmental feedback by verifying the MAC network's output against a target proxy and leverages a novel textual backpropagation to update the network.
To extend coding capabilities beyond function-level tasks to more challenging software-level development, we further propose rSDE-Bench, a requirement-oriented software development benchmark, which features complex and diverse software requirements along with automatic evaluation of requirement correctness.
Our experiments show that:
i) The automatic requirement-aware evaluation in rSDE-Bench closely aligns with human evaluations, validating its reliability as a software-level coding benchmark.
ii) EvoMAC outperforms previous SOTA methods on both the software-level rSDE-Bench and the function-level HumanEval benchmarks, reflecting its superior coding capabilities. The benchmark can be downloaded at \href{https://yuzhu-cai.github.io/rSDE-Bench/}{https://yuzhu-cai.github.io/rSDE-Bench/}.
\end{abstract}
\vspace{-4mm}
\section{Introduction}
\vspace{-2mm}

Automatic software development focuses on generating code from natural language requirements. 
Code is a universal problem-solving tool, and this automation presents significant potential to provide substantial benefits across all areas of our lives~\cite{Li2022CompetitionlevelCG}.
Recently, the industry has introduced several large language model (LLM)-driven coding assistants, including Microsoft's Copilot~\cite{Copilot}, Amazon's CodeWhisperer~\cite{CodeWhisperer}, and Google's Codey~\cite{Codey}. These coding assistants significantly advance human efficiency and yield considerable commercial benefits. Despite the initial success of LLMs in assisting with line-level coding, they struggle to tackle more complex coding tasks. This limitation stems from the restricted reasoning abilities of single LLMs and their lack of capacity for long-context understanding~\cite{wang2024adaleval,li2024looglelongcontextlanguagemodels,wang2024limitssurveytechniquesextend}.

To handle function-level coding tasks, numerous multiple language agent collaboration (MAC) systems have been proposed~\cite{li2023camel,metagpt2023,chanchateval,mapcoder2024,Yang2024SWEagentAI,Li_2022,GPTEngineer}. These MAC systems function as LLM-driven agentic workflow.
They follow human-designed standardized operating procedures to divide the complex coding tasks into simpler subtasks within the workflow, allowing each agent to conquer specific subtasks.
These MAC systems significantly advance coding capabilities from line-level to function-level tasks.  However, current MAC systems rely on heuristic designs. 
These human-crafted static systems have two inherent limitations: 
i) their performance is confined to human initialization. Given the diversity of real-world coding tasks, human design cannot fully address the specific needs of each task; and
ii) they lack the flexibility to adapt to new tasks. This rigidity necessitates that researchers and developers manually decompose tasks and create prompts. The complexity of this process inhibits effective human optimization for adapting to new challenges.

To address these limitations, we present EvoMAC, a novel self-evolving paradigm for MAC networks. EvoMAC's key feature is its ability to iteratively adapt both agents and their connections during test time for each task. 
Inspired from the standard neural network training, the core idea of self-evolution is to obtain text-based environmental feedback by verifying the MAC network's generation against a target proxy, then leverage a novel textual back-propagation to update the MAC network. Following this general paradigm, we specify EvoMAC for software development, which comprises three essential components: i) an adaptable MAC network-based coding team that generates code through feed-forward; ii) a specifically designed testing team that creates unit test cases serving as the target proxy and verifies the generated code in the compiler to produce objective feedback; and iii) an updating team that uses the textual back-propagation algorithm to update the coding team. By cycling these three components, the coding team can iteratively evolve and generate codes that are better aligned with the unit test cases, eventually fulfilling more  requirements of the coding task.

Our self-evolving MAC network has the potential to further advance coding capabilities from function-level to more complex software-level tasks. As it can iteratively address lengthier task requirements and cater to realistic software development demands. However, existing benchmarks typically focus on specific individual functions~\cite{humaneval2021,mbpp2021,CodeScore,xcodeeval} or bug-fixing~\cite{swebench2023}, leaving a significant gap in providing comprehensive requirements for software development. This gap makes it difficult to fully assess the potential of our self-evolving MAC network.

To support the development of software-level coding capabilities, we propose rSDE-Bench, a novel requirement-oriented software development benchmark. 
It is the first benchmark that features both complex and diverse software requirements, as well as the automatic evaluation of requirement correctness. rSDE-Bench involves 53 coding tasks with 616 requirements, covering two typical software types, Website, and Game, and two requirement difficulty levels, Basic and Advanced.  Each coding task consists of two components: i) multiple requirements that clearly outline measurable software functionalities, item by item, and ii) paired black-box test cases that automatically verify the correctness of each requirement. 
rSDE-Bench can achieve automatic evaluation with these synchronized pairs of requirements and test cases.
The rSDE-Bench introduces new software-level challenges, including lengthy requirement analysis and long-context coding, which are essential in real-world software development but are absent in existing benchmarks.

To validate the effectiveness of our proposed EvoMAC and rSDE-Bench, we conduct three key evaluations. 
First, we compare our automatic evaluation in rSDE-Bench with human evaluation, achieving a coherence score of 99.22\%, demonstrating its reliability. 
Second, we compare EvoMAC against five multi-agent and three single-agent baselines. EvoMAC significantly outperforms previous SOTAs by 26.48\%, 34.78\%, and 6.10\% on Website Basic, Game Basic, and HumanEval, respectively, underscoring its effectiveness. Third, we evaluate EvoMAC with varying evolving times and two different driving LLMs. The results indicate that EvoMAC consistently improves with more evolving times and shows convincing enhancements regardless of the driving LLM used, further demonstrating the effectiveness of our self-evolving design.

To sum up, our contributions are:

$\bullet$ We propose EvoMAC, a novel self-evolving MAC network, and apply it to software development. EvoMAC can iteratively adapt both agents and their connections during test time for each task. 

$\bullet$ We propose rSDE-Bench, a novel requirement-oriented software development benchmark. It is the first benchmark that features both complex and diverse software requirements, as well as the automatic evaluation of requirement correctness.

$\bullet$ We conduct comprehensive experiments and validate that: automatic evaluation in rSDE-Bench is highly aligned with human evaluation; EvoMAC outperforms previous SOTAs, and self-evolving promises continuous improvement with evolving times.

\vspace{-4mm}
\section{Related Works}
\vspace{-3mm}
\noindent\textbf{LLM-based multi-agent collaboration.} LLM-driven multi-agent collaboration (MAC) systems~\cite{werewolf2023, waragent2023, ziems2024can, autogen2023, metagpt2023, chanchateval, mandi2024roco} enable multiple agents to share information and collaboratively complete the overall task. These MAC systems function as agentic workflows. They have demonstrated enhanced problem-solving capabilities in various domains, such as mathematics~\cite{mapcoder2024}, software development~\cite{chatdev2023, metagpt2023}, embodied task~\cite{roco2024} and social simulation~\cite{ziems2024can, matrix2024, agenthospital}. However, these systems~\cite{autogen2023, agentverse2023} heavily rely on manually designed workflows, which lack generalizability and the labor-intensive nature of manual design poses significant limitations. To address this issue, we propose a novel self-evolving paradigm, which allows agents to update and improve through external feedback, enabling dynamic adaptation and more advanced performance across varied tasks.

\noindent\textbf{Software development benchmarks.} Software development benchmarks aim to evaluate models in the task of generating code from natural language descriptions~\cite{Zheng2023TowardsAU}. These benchmarks typically include task definitions and evaluation criteria. Existing benchmarks can be categorized into three types: i) function completion (HumanEval~\cite{humaneval2021}, MBPP~\cite{mbpp2021}, EvalPlus~\cite{evalplus}, xCodeEval~\cite{xcodeeval}); ii) bug repair (SWE-bench~\cite{swebench2023}); and iii) software generation (SRDD~\cite{chatdev2023}, SoftwareDev~\cite{metagpt2023}).
Function completion and bug repair benchmarks are confined to function-level task definitions, missing the diverse realistic software requirements. Software generation benchmarks often depend on expensive human evaluations or indirect similarity-based measurements, unable to automatically and accurately verify the requirement correctness.
To address these limitations, we introduce rSDE-Bench, the first benchmark contains both diverse software requirements and automatic evaluation of requirement correctness. It can support the development of more realistic software-level coding capabilities.

\vspace{-4mm}
\section{EvoMAC: Self-Evolving Multi-Agent Collaboration Network}
\vspace{-2mm}

This section presents~\texttt{EvoMAC}, a novel self-evolving multi-agent collaboration network and its application to software development. The key feature of~\texttt{EvoMAC} is its ability to iteratively adapt both agents and their connections during test-time for each task, mimicking the back-propagation process, a core algorithm in neural network training. We first formulate a general self-evolving paradigm in Sec.~\ref{System:framework} and then describe its application to software development in Sec.~\ref{System:evolving}. 

\vspace{-2mm}
\subsection{A general self-evolving paradigm via textual backpropagation}\label{System:framework}
\vspace{-2mm}

\noindent\textbf{Multi-agent collaboration network.} A multi-agent collaboration (MAC) network is a computational graph representing agentic workflows, where multiple agents empowered by LLMs interact as interconnected nodes to coordinate and share information for complex task-solving. The intuition behind to divide the complex task into more specific and manageable subtasks for each agent, allowing the overall task to be gradually conquered through the agentic workflow. Mathematically, we represent a MAC network with $N$ autonomous agents as a directed acyclic graph $\mathcal{A} = (\mathcal{V}, \mathcal{E})$, where $\mathcal{V} = \{v_i\}_{i=1}^{N}$ is the set of $N$ nodes, and $\mathcal{E} = \{e_{i,j}\}_{i,j \in [1, \dots, N], i \neq j}$ is the set of directed edges with no circles. The $i$-th node $v_{i}$ represents the $i$-th autonomous agent with the prompt $p_{i}$, which specifies its subtask. The edge $e_{i,j}$ represents the task dependency between the $i$-th agent and the $j$-th agent, indicating that the $j$-th agent's subtask should be executed after the $i$-th agent's subtask in the agentic workflow. The overall graph topology specifies the agentic workflow. Analogy to traditional neural networks, agents function similarly to neurons, with agent prompts serving as neurons' weights and the agentic workflow as the layers and connections.

The feed-forward pass of MAC network is the execution of the agentic workflow. In this process, each agent is given two inputs: the initial task requirement and the output from the previous agent. Using these, each agent produces an output that fulfills its specific subtask. Eventually, the last agent's generation constitutes the final output, integrating all completed subtasks. Note that the initial task requirement is input to each agent as context, providing supplementary details to aid in the implementation of each subtask.


Recently, various MAC networks have been designed using human expertise to assign fixed agent prompts and workflows~\cite{metagpt2023, chanchateval}, resembling untrained neural networks. However, these designs solely rely on human priors and lack adaptability, causing limited performance improvement over a single agent. To overcome this, inspired by neural network training, we propose a self-evolving paradigm for multi-agent collaboration networks, enabling both agents and their connections to dynamically evolve during test-time for each given task.




\begin{figure}
    \centering
    \includegraphics[width=1.0\linewidth]{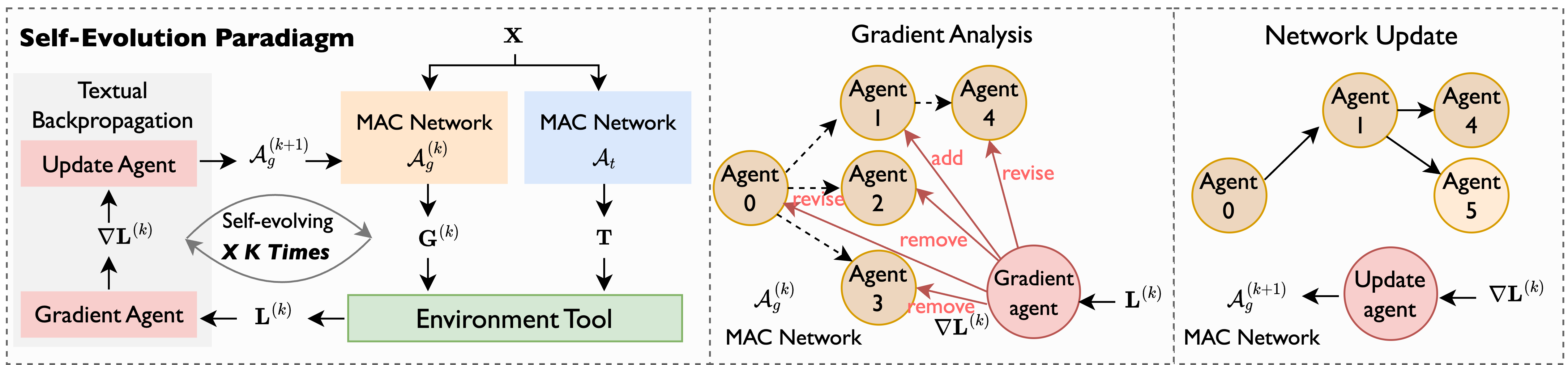}
    \vspace{-6mm}
    \caption{The general self-evolving paradigm.}
    \label{fig:framework}
    \vspace{-6mm}
\end{figure}

\noindent\textbf{Optimization problem.}
Here we consider a general generation task. During test-time, given a task, the MAC network performs a feed-forward pass to generate the final output without knowing its quality. The key to evolution during test-time is to set up a target proxy for the MAC network to guide its improvements in the generated output. Here we consider this target proxy as the conditions for task completion, such as unit tests in coding, and we can produce such a target proxy by another group of autonomous agents based on the same task description. Then, the quality of each generated output can be verified according to the target proxy. This approach relies on two key assumptions: (i) generating a target proxy is significantly simpler than completing the original generation task, and (ii) the generated output can be correctly verified against the target proxy through an objective environment. These assumptions are practical in many applications. For example, in code generation, producing unit tests, the expected input-output pairs, is much easier than generating the entire code; meanwhile, a code compiler naturally acts as the objective environment to check the correctness of the generated code against the unit test, providing objective and informative feedback.

Mathematically, let $\mathbf{X}$ be the textual description of a task. Given the MAC network $\mathcal{A}_g$, the generated output is $\mathbf{G} = \Phi(\mathbf{X}, \mathcal{A}_g)$, where  $\Phi(\cdot, \cdot)$ is the general feed-forward operator that executes the agentic workflow, processing the input text through the MAC network. Similarly, the target proxy is $\mathbf{T} = \Phi(\mathbf{X}, \mathcal{A}_t)$, where $\mathcal{A}_t$ is another MAC network designed for producing the target proxy. Note that we aim to evolve and optimize $\mathcal{A}_g$, while keep $\mathcal{A}_t$ predefined and fixed. The optimization of our self-evolution is formulated as, 
\begin{equation}
\label{eq:optimization}
     \mathcal{A}_g^{*} = \underset{\mathcal{A}_g}{\min}~\langle \Phi(\mathbf{X}, \mathcal{A}_g), \mathbf{T}\rangle_{E}, ~~\text{subject~to:}~~ \mathbf{T}=\Phi\left(\mathbf{X}, \mathcal{A}_t \right),
\end{equation}
where $\langle \cdot,\cdot\rangle_{E}$ is an objective environment executor that receives the generated output and the target proxy as inputs and outputs a text-based environmental feedback.  Akin to the loss function in traditional neural network training, which quantifies the difference between the generated output and the ground-truth, the objective in~\eqref{eq:optimization} evaluates whether the generated output meets the conditions of the task completion using the environment, subsequently producing execution reports as the text-based environmental feedback. Here the minimization operation $\min$ is defined to reduce the failures during execution. With the guidance of the target proxy and the objective feedback given by the environment, the MAC network can improve its success rate of task completion during test time. 

Note that, another straightforward way to enable the MAC network's evolution is through the self-critique strategy~\cite{symbolic2024,Valmeekam2023CanLL,Xu2024ChatGLMMathIM,Asai2023SelfRAGLT}, which employs a critique agent to assess the generated output directly. This approach has two inherent limitations: i) the critique may be subjective and biased, and ii) the critique agent can have hallucinations, causing inconsistencies and errors. These limitations can cause the MAC network to become entrenched in its own preferences or evolve in the wrong direction, especially iterating multiple times; see our experimental validations in Tab.~\ref{Table:ablation study}. In comparison, our approach leverages an environment executor to provide objective feedback, preventing bias and hallucinations.

While we use the analogy between our self-evolution process and neural network training for motivating, they are significantly different in three key aspects: (i) our self-evolution occurs at test time without a dedicated training phase; (ii) it evolves for each specific task individually rather than over a batch of samples; and (iii) the environmental feedback are usually texts, not be numerical values, which cannot be optimized by the standard backpropagation. This motivates us to propose our textual backpropagation.

\noindent\textbf{Solution based on textual backpropagation.} The self-evolution solution iteratively updates the MAC network using a textual backpropagation algorithm, guided by the environmental feedback. The core idea is to analyze the influence of each agent in the MAC network $\mathcal{A}_g$ to the final environmental feedback and use these analyses to update the agent prompts and the agentic workflow in $\mathcal{A}_g$. This is achieved by two collaborative agents, each responsible for one of the two key steps: (i) textual gradient analysis and (ii) network update. The overall algorithm can refer to Alg.~\ref{alg:EvoMAC} in the appendix.

First, the gradient agent takes the environmental feedback as the input and outputs textual gradients that describe the impact of each agent in the MAC network. Let $\mathcal{A}_g^{(k)}$ and $\mathbf{L}^{(k)}$ be the MAC network and  the environment feedback at the $k$-th iteration. The textual gradient is then $\nabla \mathbf{L}^{(k)} = \mathcal{G}(\mathcal{A}_g^{(k)},\mathbf{L}^{(k)})$, where $\mathcal{G}(\cdot,\cdot)$ is the gradient analysis operator managed by the gradient agent; see its prompt in Appendix. The textual gradient details three-fold information for each agent inside $\mathcal{A}_g^{(k)}$: 1) whether this agent's subtask is fulfilled; 2) whether this agent introduces errors; and 3) whether any subtask is missed in the current MAC network.

Second, based on the textual gradients, the updating agent iterates the MAC network as $\mathcal{A}_g^{(k+1)} = \mathcal{U}(\mathcal{A}_g^{(k)}, \nabla \mathbf{L}^{(k)})$, where $\mathcal{U}(\cdot,\cdot)$ is the updating operator managed by the updating agent. This operator guides the updates from three-folds: 1) removing the agents whose subtasks have been completed; 2) revising the erroneous agent's prompts by adding potential solutions provided in the gradient analysis; and 3) adding new agents for missing subtasks and restructuring the workflows based on the subtask dependencies noted in the gradient analysis; see the prompt details in Appendix. These adjustments address existing issues and fulfill unmet requirements in the current generation of the MAC network, promising improvements in the updated version.


Note that, the key of the textual backpropagation is the prompt designs for both gradient analysis and network updates. The design must i) thoroughly evaluate the subtask of each agent in the MAC network according to the objective environment feedback and determine necessary adjustments to the MAC network to address existing issues, fulfilling the unmet requirements; and ii) maintain coherence, ensuring that issues identified by the gradient agent can be effectively resolved by the updating agent's modifications to the MAC network.

\begin{figure}
    \centering
    \includegraphics[width=1.0\linewidth]{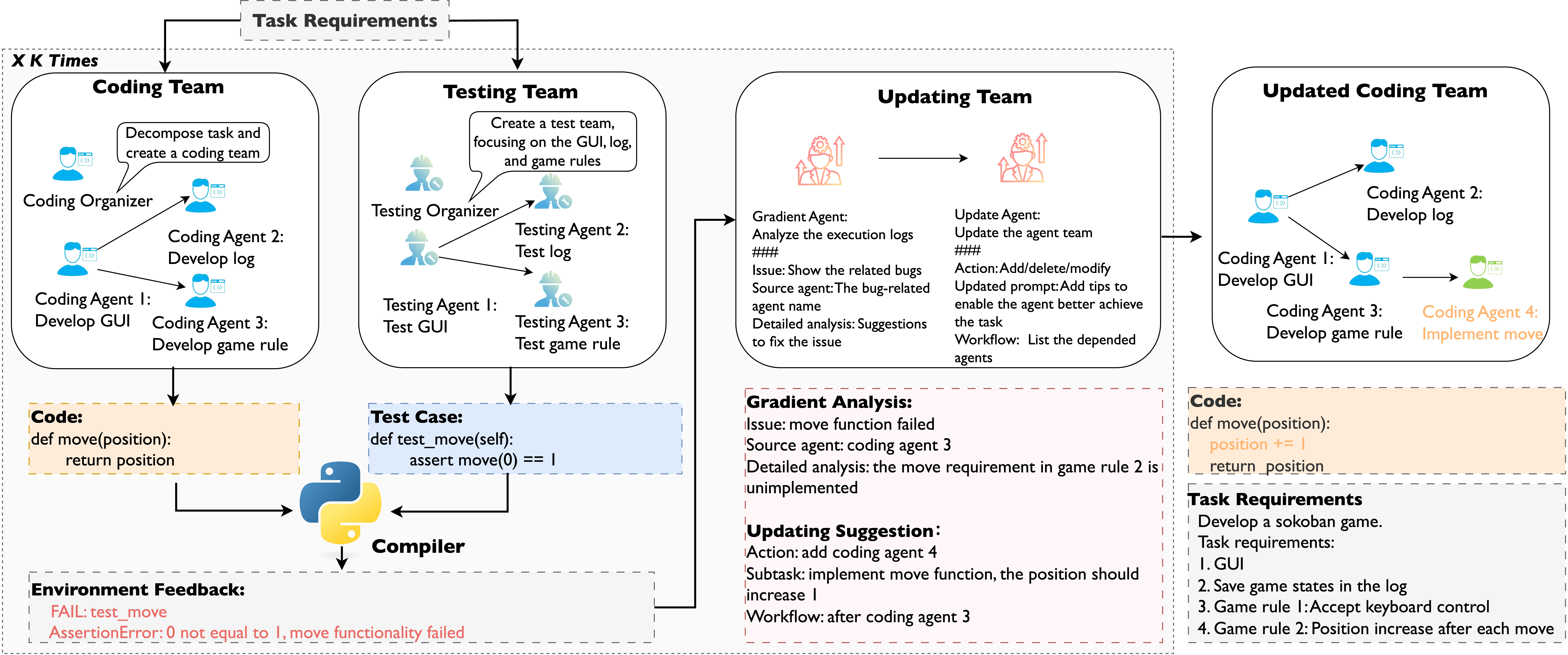}
    \vspace{-6mm}
    \caption{EvoMAC takes task requirements as input and iteratively updates the coding team to generate code that better fulfills the requirements.}
    \label{fig:system}
    \vspace{-6mm}
\end{figure}

\vspace{-2mm}
\subsection{Self-evolution for software development}
\label{System:evolving}
\vspace{-2mm}

In this section, we apply the self-evolving paradigm to the task of software development. The overall architecture of the proposed self-evolving multi-agent collaboration network for software development is illustrated in Fig.~\ref{fig:framework}. Given a coding task, the coding team, corresponding to the MAC network $\mathcal{A}_g$, generates all the codes through its forward-pass; the testing team, associated with the MAC network $\mathcal{A}_t$, is responsible for creating the target proxy; that is, unit tests of the coding task; and the objective environment tool is realized through the compiler. The identified bugs during execution form the textual environmental feedback. The updating team, consisting of two collaborative agents, manages the textual backpropagation. By continuously cycling through feed-forward, feedback collection, and textual backpropagation processes, the coding team is iteratively refined to more closely align with the test cases. The detailed implementation of agents can refer to Sec.~\ref{app:prompts} in the Appendix.

Since unit test generation is much easier than the original logical code generation, the testing team usually can produce high-quality test cases, which are closely aligned with the task requirements. Then, improving alignment with the unit tests through MAC network updates ensures better adherence to the actual task requirements.

\noindent\textbf{Coding team for feed-forward.} In the feed-forward process, the coding team synthesizes code according to the given coding task. To handle the extensive software requirements, the coding team is implemented as a MAC network. It divides the comprehensive requirements into a sequence of smaller, more specific function implementation subtasks, and progressively conquers them through the agentic workflow. 
Unlike existing MAC systems that heuristically decompose coding tasks and define the agentic workflow, we initialize the MAC network using a novel self-organizing approach.
A coding organizer agent automatically and flexibly decomposes the task requirements into subtasks and assembles the coding agent team accordingly. The number of coding agents is dynamic, adjusting in response to the task requirements. Note that, the quality of the generated code is unknown during the forward pass, which necessitates the self-evolving paradigm to iteratively refine the generation.

\noindent\textbf{Testing team and compiler for feedback collection.} 
To verify whether the generated code meets the requirements of the coding task, we employ unit tests as the target proxy. These test cases consist of input-output pairs tailored to specific requirements. For example, a test case for a keyboard control requirement would detail the type of control as the input and describe the expected behavior as the output. To create flexible and comprehensive unit tests, we set up the testing team as a MAC network and also initialize it in a self-organized way. A testing organizer agent automatically decomposes our specified key testing criteria into subtasks and accordingly forms the testing agent team .


Once the test cases and generated code are ready, they are executed in the compiler, which functions as the environmental tool, producing execution logs. These logs clearly point out the gap between the generated code and the test cases. It shows satisfied testing requirements, existing function errors, and unmet testing requirements. This feedback information can be used to verify whether each agent's subtask is accomplished and guide the MAC network update.

\noindent\textbf{Updating team for textual back-propagation.}
The updating team consists of two collaborative agents: the gradient agent and the updating agent, adjusting the MAC network based on the execution logs, including the agent prompts and workflows. This process consists of two steps. First, the gradient agent summarizes the textual gradient by identifying accomplished subtasks for satisfied requirements, appending new subtasks for unmet requirements, and analyzing errors to detail their originating agents and revising suggestions. Second, the updating agent modifies the coding agent team by removing agents that have completed their subtasks, adding new agents for the new subtasks, and revising agent prompts to address issues identified in the previous generation. The agent workflow is updated once the agent team is revised, based on the dependencies among the subtasks.

\vspace{-4mm}
\section{rSDE-Bench: requirement-Oriented Software Development Engineering Benchmark}
\vspace{-2mm}

This section introduces \texttt{rSDE-Bench}, a requirement-oriented benchmark designed to assess the ability of models to handle software-level coding tasks. Each coding task involves multiple detailed software requirements. These requirements specify each functionality and constraint of the software, item by item, serving as measurable benchmarks for assessing the software’s effectiveness.
As shown in Fig.~\ref{fig:evalation}, unlike previous instruction-oriented approaches~\cite{chatdev2023,metagpt2023} which rely on brief instructions as input, \texttt{rSDE-Bench} uses comprehensive software requirements as input, complemented by unit test cases to automatically evaluate the correctness. This benchmark provides software-level coding tasks and automatic evaluation, aligning more closely with real-world software development practices.


\vspace{-4mm}
\subsection{Benchmark construction}
\label{Benchmark:construction} 
\vspace{-2mm}



rSDE-Bench involves two typical real-world software types: game and website. They can reflect different coding capacities demanded in realistic software development.
Game often requires handling dynamic interactions, real-time state changes, and user-driven operations, focusing on elements like logic execution, initialization, and game state transitions. Website emphasizes static and dynamic content management, user interaction through forms and buttons, and ensuring page elements are displayed and functional. 
rSDE-Bench involves diverse requirements, each paired with a test case. Specifically, rSDE-Bench provides 53 unique coding tasks and 616 test cases. For details on the benchmark construction, software statistics, software requirements, and test case examples, see Sec.~\ref{app:benchmark} in the Appendix.

rSDE-Bench introduces two requirement difficulty levels, including basic and advanced, to reflect the varying complexity of real-world software development tasks. The basics reflect the fundamental and more achievable requirements, such as interaction, control, and logging. The advanced reflects more complex software functionalities, such as game logical rules, and dynamic web content management. The details can be referred to the appendix.

\vspace{-2mm}
\subsection{Automatic Evaluation}
\vspace{-2mm}

\begin{figure}[!t]
    \centering
    \includegraphics[width=1.0\linewidth]{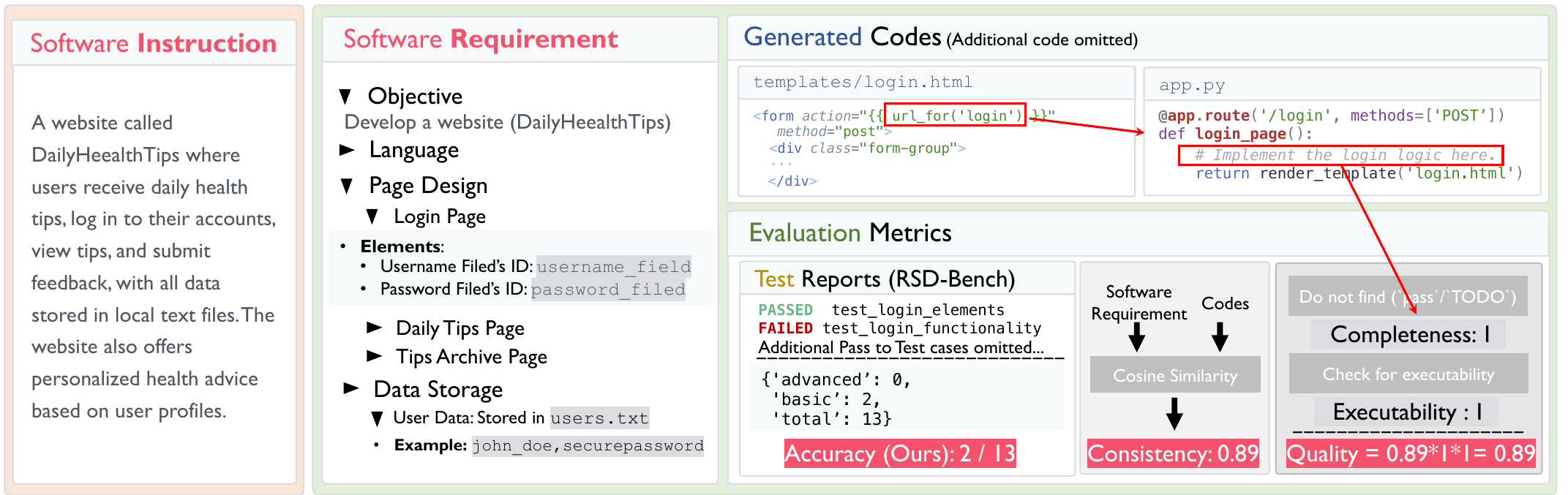}
    \vspace{-6mm}
    \caption{Comparison between instruction-oriented and requirement-oriented evaluations. rSDE-Bench accurately reflects requirement fulfillment with the proposed accuracy score of $2/13$, while the indrection evaluation misjudges with high scores ($0.89$), failing to detect missing functionality.}
    \vspace{-4mm}
    \label{fig:evalation}
\end{figure}

rSDE-Bench supports automatic evaluation of requirement correctness.
It achieves this by pairing a specifically designed black-box test case with each requirement. The test case can directly verify whether the generated code achieved the requirement.
Its evaluation metric is the accuracy, which quantifies the proportion of correctly passed test cases. It is similar to the \texttt{pass@1} metric in HumanEval~\cite{humaneval2021}, which evaluates the pass ratio of correctly achieved functions against the total functions via unit test verification. 
It is a fully automated evaluation process, eliminating the need for human involvement while still providing accurate and reliable assessments.



Previous benchmarks for software code generation mainly rely on two evaluation methods. One method is human evaluation~\cite{metagpt2023}, which is time-consuming and not scalable for large datasets.
The other method is indirect evaluations~\cite{chatdev2023}, which defines metrics like consistency, completeness, and quality. Consistency measures how closely the generated software aligns with the original requirement description by comparing the cosine similarity between the two. Completeness is determined by detecting the presence of placeholder (such as \texttt{pass} or \texttt{TODO}), which results in a binary value of 0 or 1. Quality is then calculated as the product of several factors: consistency, completeness, and executability.
As illustrated in Fig.~\ref{fig:evalation}, they could not measure the correctness of the generated code in fulfilling requirements. 
In contrast, \texttt{rSDE-Bench}'s test cases-based evaluation is more rigorous and precise. These test cases can accurately verify the correctness of generated code in fulfilling the requirements.
\texttt{rSDE-Bench} promises reliable and scalable automatic evaluation. In the experiments, we have validated the significant advantages of the proposed automatic evaluation over the previous metrics, including consistency and quality; see Fig.~\ref{Table:humanevalpercentage}.


\vspace{-2mm}
\subsection{Features}
\label{Benchmark:features}
\vspace{-2mm}


\noindent\textbf{Challenging and diverse software requirements.} rSDE-Bench features long-context software requirements (averaging 507/1011 words for game and website tasks, respectively), unlike instruction-oriented benchmarks~\cite{humaneval2021,mbpp2021,swebench2023} that rely on brief prompts. These detailed requirements better reflect real-world lengthy and complex software development challenges.

\noindent\textbf{Requirement-aware precise and efficient evaluation.} 
rSDE-Bench employs detailed software requirements and automated unit tests to precisely measure how well generated software meets its objectives. Generated codes are evaluated based on pass rates from running specific test cases, offering an accurate and efficient process. In contrast, instruction-oriented benchmarks rely on brief prompts, which lack constraints and make evaluation less reliable, often requiring labor-intensive or indirect evaluation.

\vspace{-3mm}
\section{Experiments}
\vspace{-3mm}

\subsection{Experimental setup}
\vspace{-2mm}

\noindent\textbf{Baselines.} To validate the effectiveness of our EvoMAC, we conducted comparisons against both single-agent and multi-agent baselines. The single-agent baselines involve three prominent large models: GPT-4o-Mini (gpt-4o-mini), Claude-3.5-Sonnet (claude-3-5-sonnet-20240620), and Gemini (gemini-1.5-flash). For multi-agent baselines, we included five state-of-the-art (SOTA) methods: MetaGPT~\cite{metagpt2023}, Autogen~\cite{autogen2023}, Mapcoder~\cite{mapcoder2024}, Agentverse~\cite{agentverse2023}, and ChatDev~\cite{chatdev2023}. To ensure a fair comparison, all multi-agent baselines, including our EvoMAC, are powered by the efficient and powerful GPT-4o-Mini model. Additionally, to demonstrate the adaptability and robustness of our EvoMAC, we developed two EvoMAC variants using GPT-4o-Mini and Claude-3.5-Sonnet.


\noindent\textbf{Datasets.} Our experiments cover both the proposed rSDE-Bench and the standard coding benchmark HumanEval~\cite{humaneval2021}. HumanEval comprises 164 Python function completion problems, where the task is to generate code from a single function description. 


\begin{figure}[!t]
    \centering
    \includegraphics[width=1.0\linewidth]{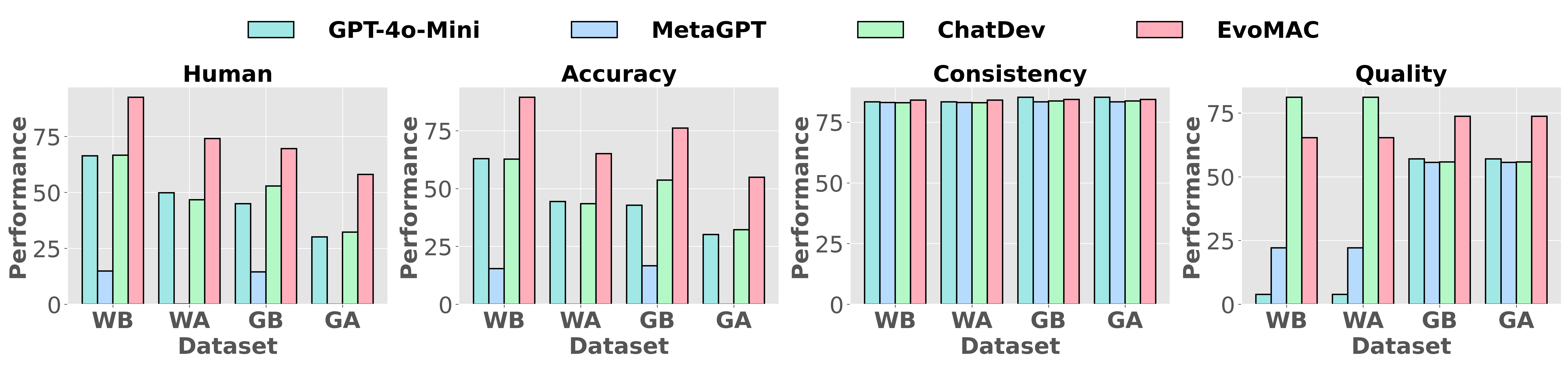}
    \vspace{-8mm}
\caption{Performance of four methods in terms of four evaluation metrics, including human evaluation, our automatic evaluation (accuracy), consistency, and quality. WB/GB and WA/GA represent Web/Game Basic and Web/Game Advanced respectively. Our accuracy metric is highly aligned with human evaluation across four dataset settings.} 
\label{Table:humanevalpercentage}
\vspace{-2mm}
\end{figure}
\begin{table}[!t]
\vspace{-2mm}
\setlength{\arrayrulewidth}{0.05pt} 
\caption{Comparison of EvoMAC with five multi-agent and three single-agent SOTA baselines, all powered by GPT-4o-Mini. \textbf{\color{red}{Red}} values represent the percentage improvement of EvoMAC, shade in pink, over the single-agent baselines, shade in grey.} 
\vspace{-3mm}
\centering
\scalebox{0.85}{
\begin{tabular}{c|c|cccc|c}
\hline

 &
   &
  \multicolumn{4}{c|}{\textbf{rSDE-Bench}} &
  \multirow{2}{*}{\textbf{\begin{tabular}[c]{@{}c@{}}HumanEval\\ (\%)\end{tabular}}} \\

 &
   &
  \multicolumn{2}{c}{\textbf{Website(\%)}} &
  \multicolumn{2}{c|}{\textbf{Game(\%)}} &
   \\ 

\multirow{-3}{*}{\textbf{Method}} &
  \multirow{-3}{*}{\textbf{Model}} &
  Basic &
  \multicolumn{1}{c}{Advanced} &
  Basic &
  Advanced &
  Pass@1 \\ \hline
 &
  Gemini-1.5-Flash &
  29.79±1.00&
  \multicolumn{1}{c|}{11.61±2.34} &
  21.74±6.39 &
  6.45±6.97 &
  73.17 \\
 &
  Claude-3.5-Sonnet &
  58.90±1.48 &
  \multicolumn{1}{c|}{37.11±1.06} &
  44.20±5.41 &
  18.29±13.26 &
  89.02 \\

\multirow{-3}{*}{Single-Agent} &
  \cellcolor[HTML]{e6e6e6} GPT-4o-Mini &
   \cellcolor[HTML]{e6e6e6}62.90±2.52&
  \multicolumn{1}{c|}{\cellcolor[HTML]{e6e6e6}44.40±4.21} &
  \cellcolor[HTML]{e6e6e6}42.76±15.50 &
  \cellcolor[HTML]{e6e6e6}30.10±11.87 &
  \cellcolor[HTML]{e6e6e6}88.41 \\ \hline
 &
  MetaGPT &
   15.41±0.00 &
  \multicolumn{1}{c|}{0.00±0.00} &
  16.67±2.71 &
  0.00±0.00 &
  88.41 \\
 &
  Autogen &
  25.68±4.14 &
  \multicolumn{1}{c|}{5.40±3.34} &
  17.39±1.78 &
  0.00±0.00 &
  85.36 \\
 &
  MapCoder &
  34.70±1.59 &
  \multicolumn{1}{c|}{14.57±0.66} &
  29.71±6.72 &
  7.52±6.10 &
  90.85 \\
 &
  Agentverse &
   15.41±0.00 &
  \multicolumn{1}{c|}{0.00±0.00} &
  37.67±8.20 &
  16.13±4.55 &
  90.85 \\
\multirow{-3}{*}{Multi-Agent} &
  ChatDev &
  62.67±0.28 &
  43.45±0.77 &
 \multicolumn{1}{|c}{ 53.63±5.70}  &
  \multicolumn{1}{c|}{32.26±4.55}  &
  70.73 \\ \cline{2-7}
 &\cellcolor[HTML]{FFF7F7}
 &
 \cellcolor[HTML]{FFF7F7} \textbf{89.38±1.01} &
  \multicolumn{1}{c|}{\cellcolor[HTML]{FFF7F7}\textbf{65.05±1.56}} &
  \cellcolor[HTML]{FFF7F7}\textbf{77.54±2.04} &
  \cellcolor[HTML]{FFF7F7}\textbf{51.60±4.54} &
  \cellcolor[HTML]{FFF7F7}\textbf{94.51} \\

 &
  \multirow{-2}{*}{\cellcolor[HTML]{FFF7F7}EvoMAC} &
  {\color[HTML]{FF0000}\cellcolor[HTML]{FFF7F7}\textbf{+26.48}} &
  \multicolumn{1}{c|}{\color[HTML]{FF0000}\cellcolor[HTML]{FFF7F7}\textbf{+20.65}} &
  { \color[HTML]{FF0000}\cellcolor[HTML]{FFF7F7}\textbf{+34.78}} &
  { \color[HTML]{FF0000}\cellcolor[HTML]{FFF7F7}\textbf{+21.50}} &
  { \color[HTML]{FF0000}\cellcolor[HTML]{FFF7F7}\textbf{+6.10}} \\ \hline
\end{tabular}}
\label{Table:sota game web percentage}
\label{Tabel:sota humaneval}
\vspace{-6mm}
\end{table}

\vspace{-4mm}
\subsection{Effectiveness of rSDE-Bench's evaluation and EvoMAC}
\vspace{-2mm}

\noindent\textbf{rSDE-Bench's automatic evaluation metric (accuracy) is highly aligned with human evaluation.}
Our primary goal is to validate the effectiveness of the proposed automatic evaluation in rSDE-Bench by comparing it with two existing evaluation metrics: consistency and quality, both from SRDD~\cite{chatdev2023}.
For a fair comparison, our golden standard is human evaluation, conducted by two expert code engineers who manually verify the fulfillment of requirements by interacting with the developed software. 
This process is tedious, taking around four hours per expert to evaluate the entire benchmark. The effectiveness of an evaluation metric depends on how closely it aligns with human evaluation.


Fig.~\ref{Table:humanevalpercentage} presents the performance of four methods in terms of four evaluation metrics, including human evaluation, our automatic evaluation, consistency, and quality. 
We see that: i) our automatic evaluation is highly aligned with human evaluation across two software types (Website and Game), four methods, (GPT-4o-Mini, MetaGPT, ChatDev, and our EvoMAC), and two requirement difficulties (Basic and Advanced). The correlation coefficient between human evaluation and our accuracy metric is 0.9922, demonstrating the effectiveness of the proposed automatic evaluation in rSDE-Bench; ii) Consistency and quality metrics differ significantly from human evaluation, with correlation coefficients of 0.2583 and 0.3041, respectively.
This discrepancy occurs because consistency in SRDD measures similarity, and quality in SRDD focuses on executability, which does not guarantee that all requirements are met. This highlights the need for rSDE-Bench, as the SRDD benchmark does not support requirement-oriented software development.


\noindent\textbf{EvoMAC outperforms previous SOTAs on both software-level and function-level coding benchmarks: rSDE-Bench and HumanEval.} 
Tab.~\ref{Table:sota game web percentage} compares EvoMAC with five multi-agent and three single-agent SOTA baselines, all powered by GPT-4o-Mini for a fair comparison. We see that EvoMAC significantly outperforms previous SOTAs across all datasets. EvoMAC outperforms single-agent methods by 26.48\% on the rSDE-Bench Website Basic and 34.78\% on the rSDE-Bench Game Basic, as well as surpassing existing multi-agent methods by over 20\%. This highlights the effectiveness of multi-agent collaboration and the power of EvoMAC.

\vspace{-4mm}
\subsection{Effectiveness of evolving}
\vspace{-2mm}

\begin{figure}[!t]
    \centering
    \begin{subfigure}{0.4\linewidth}
    \includegraphics[width=0.45\linewidth]{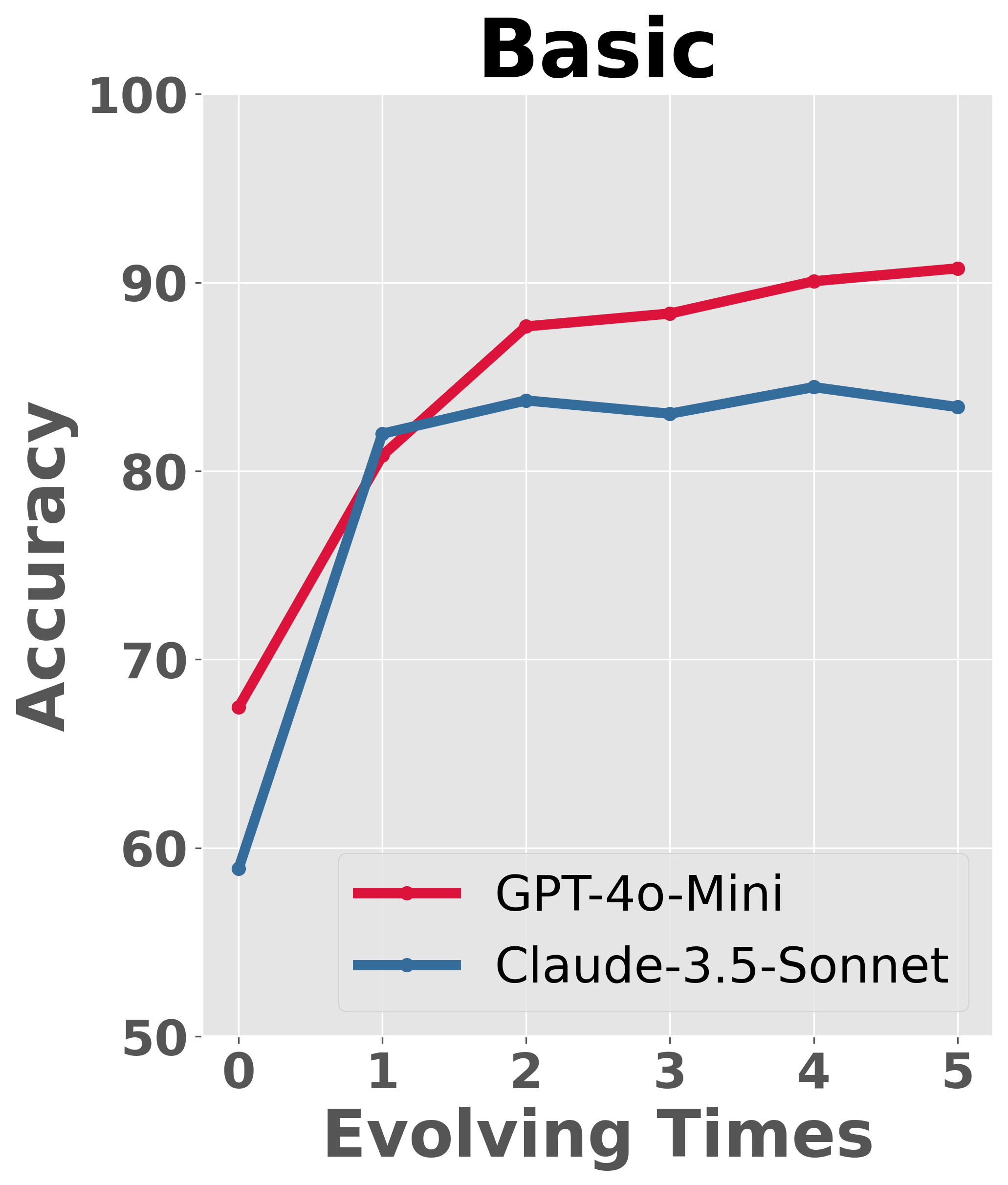}
    \includegraphics[width=0.43\linewidth]{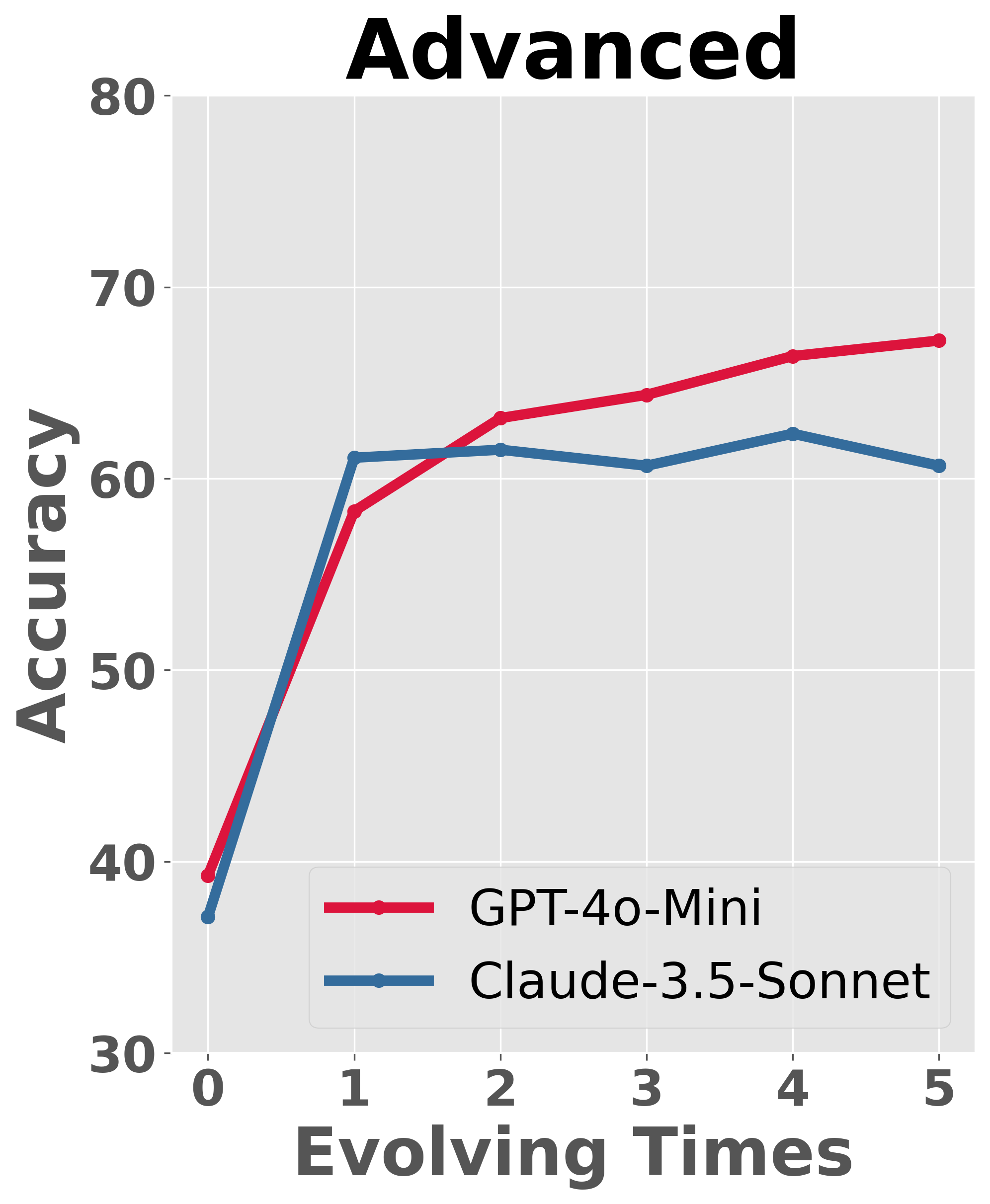}
    \caption{Website}
    \label{fig:evolving-website}
    \end{subfigure}
    \hfill
    \begin{subfigure}{0.4\linewidth}
    \includegraphics[width=0.43\linewidth]{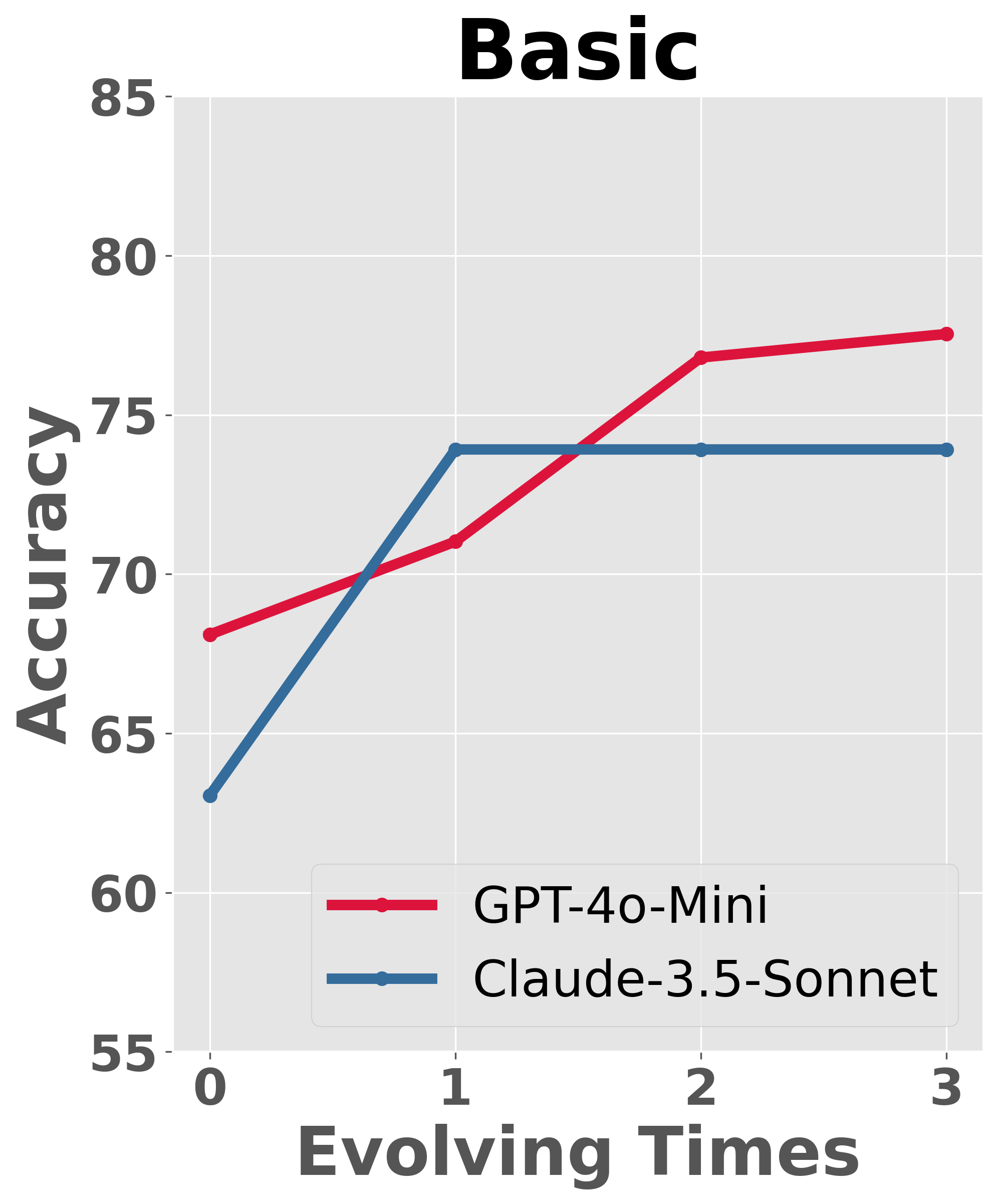}
    \includegraphics[width=0.43\linewidth]{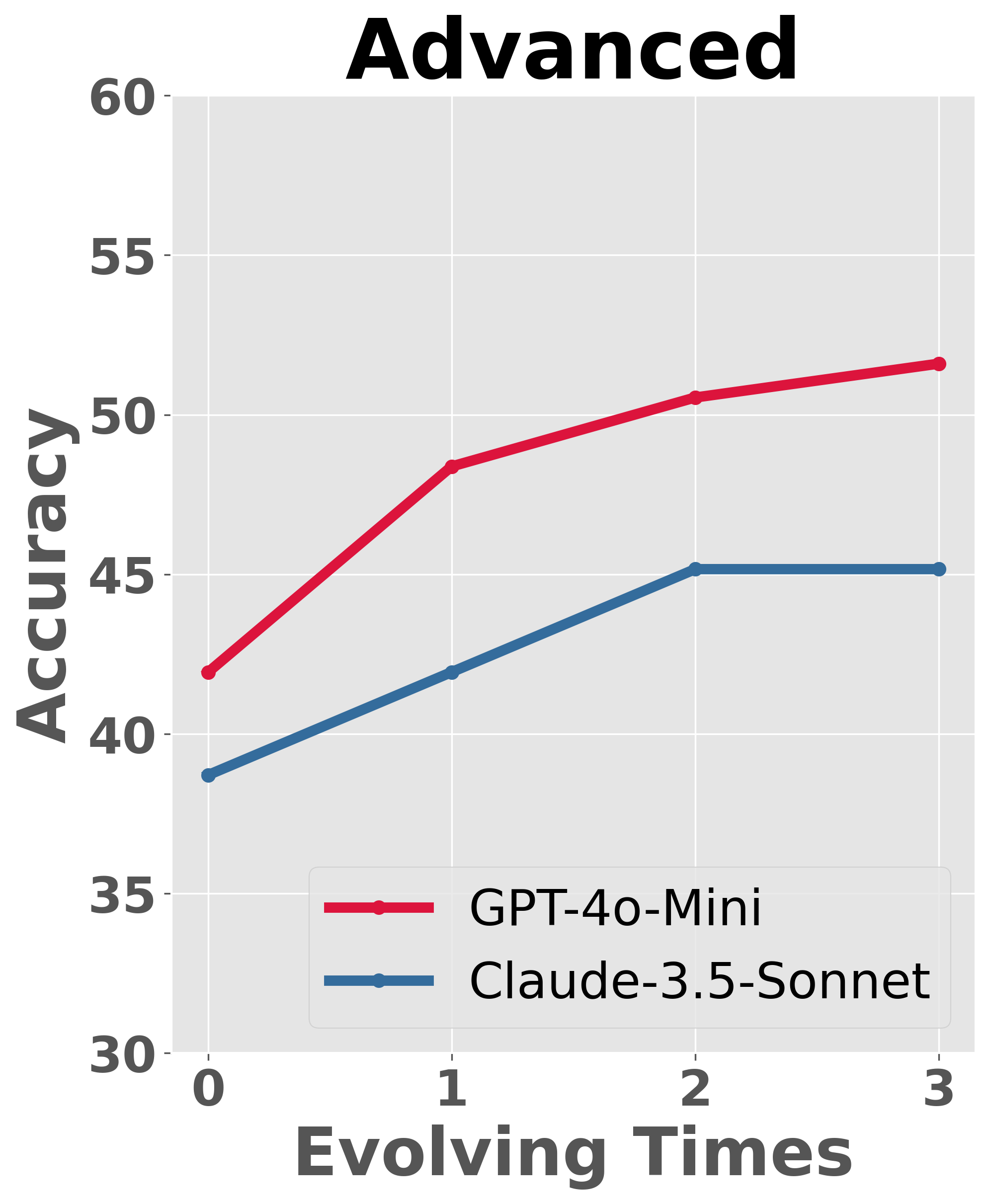}
    \caption{Game}  
    \label{fig:evolving-game}
    \end{subfigure}
    \hfill
    \begin{subfigure}{0.18\linewidth}
    \includegraphics[width=1.0\linewidth]{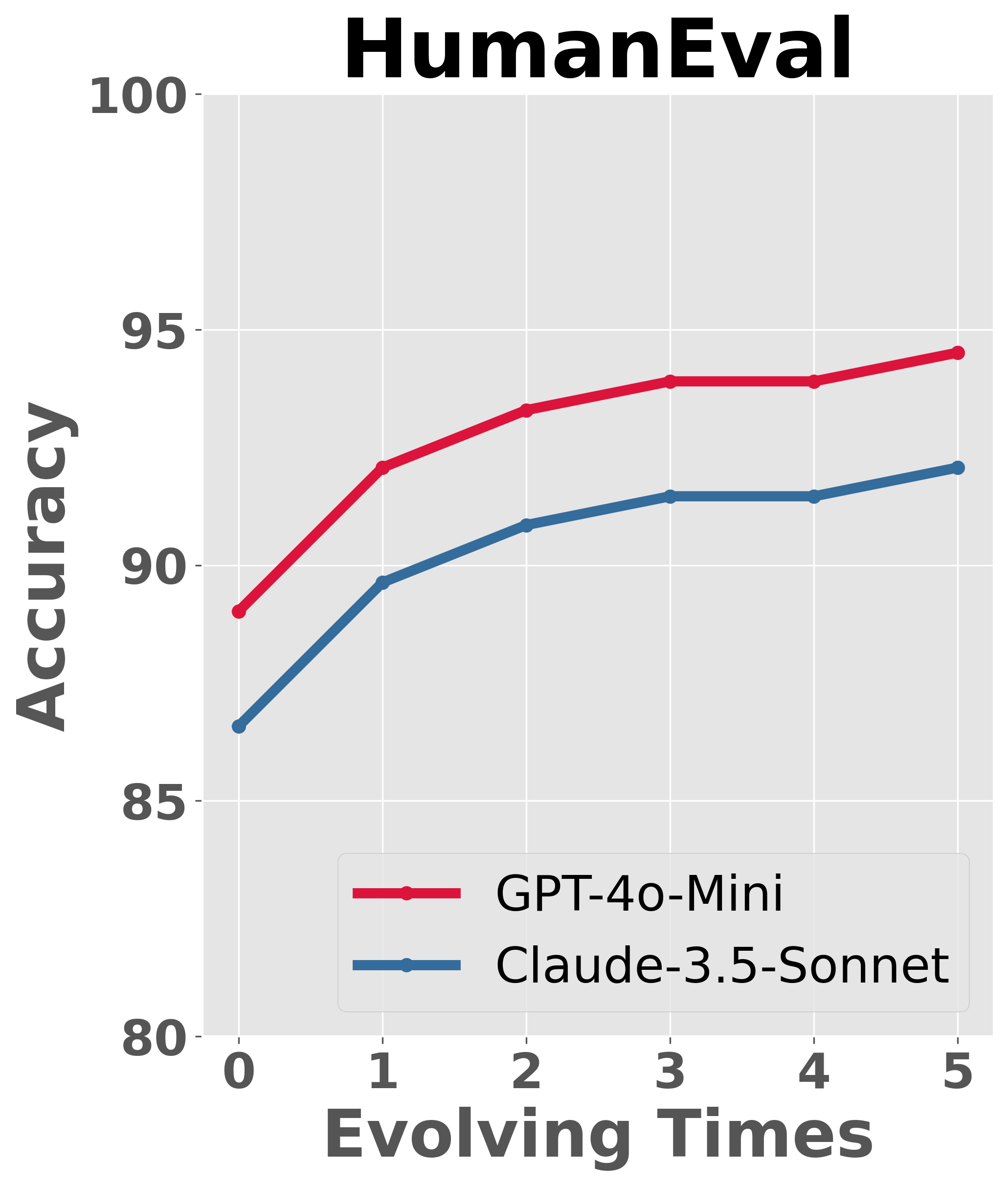}
    \caption{HumanEval} 
    \label{fig:evolving-humaneval}
    \end{subfigure}
    \vspace{-3mm}
    \caption{Effect of EvoMAC performance across evolving times empowered by GPT-4o-Mini and Claude-3.5-Sonnet on Website, Game, and HumanEval datasets. The figure shows EvoMAC continuously improves with the evolving times on both LLM drives.}
    \label{fig:evolving}
    \vspace{-6mm}
\end{figure}

Fig.~\ref{fig:evolving} shows the accuracy of EvoMAC over multiple evolving iterations on the rSDE-Bench and HumanEval. Each figure presents two curves: one for EvoMAC powered by GPT-4o-Mini (red) and the other by Claude-3.5 (blue). We have the following findings:


\begin{minipage}{0.6\textwidth}
\centering
\centering
\captionof{table}{Ablation study about coding/testing team with single/multi-agent, with/without evolving, and with/without environment tool. Best performances are bolded.}
\vspace{-2mm}
\setlength\tabcolsep{3pt}
\scalebox{0.8}{
\begin{tabular}{lcc|cc|cccc}
\toprule
\multirow{2}{*}{} & \multirow{2}{*}{Coding}                                                & \multirow{2}{*}{Testing}                                               & \multirow{2}{*}{Evol.} & \multirow{2}{*}{Env.} & \multicolumn{2}{c}{\textbf{Website(\%)}} & \multicolumn{2}{c}{\textbf{Game(\%)}} \\
                     &                                                                        &                                                                        &                           &                              & Basic             & Advanced         & Basic           & Advanced        \\ \midrule
a)                    & Single  & - & -                         & -                            & 63.70             & 41.70            & 42.76           & 30.10           \\
b)                    & Multi  & - & -                         & -                            & 67.47             & 39.27            & 68.10           & 41.93           \\
c)                    & Single  & Single  & $\checkmark$              & $\checkmark$                 & 80.82             & 60.32            & 71.73           & 41.93           \\
d)                    & Multi   & Single  & $\checkmark$              & $\checkmark$                 & 83.90             & 60.72            & 76.08           & 41.93           \\
e)                    & Single  & Multi   & $\checkmark$              & $\checkmark$                 & 83.56             & 61.94            & 73.91           & 45.16           \\
f)                    & Multi   & Multi  & $\checkmark$              & -                            & 78.08             & 52.23            & 55.80           & 33.32           \\
g)                    & Multi   & Multi   & $\checkmark$              & $\checkmark$                 & \textbf{90.75}    & \textbf{67.20}   & \textbf{77.54}  & \textbf{51.60}  \\ \bottomrule
\end{tabular}
\label{Table:ablation study}}
\vspace{-1mm}
\end{minipage}
\begin{minipage}{0.4\textwidth}
    \centering
    \includegraphics[width=1.0\linewidth]{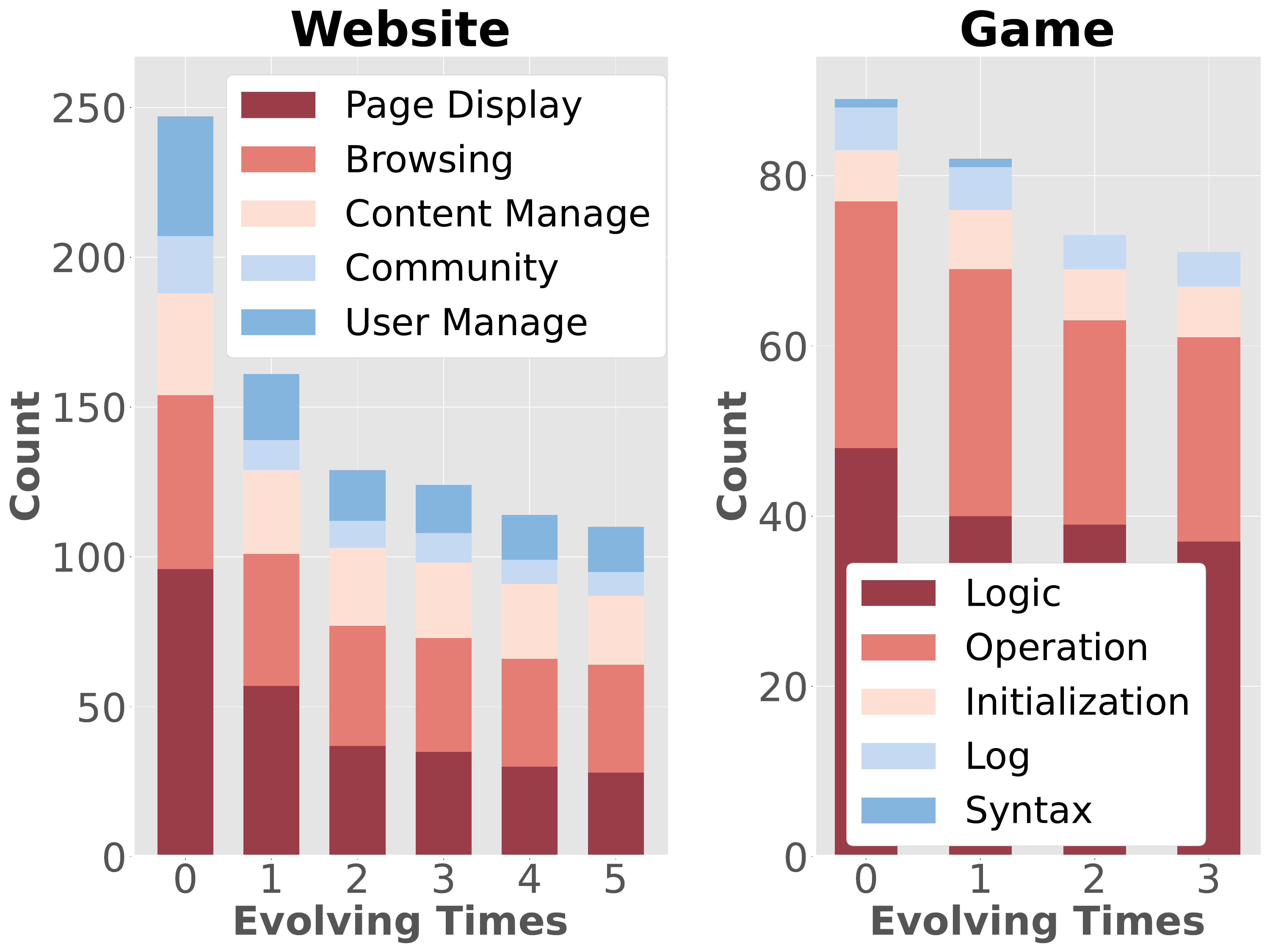}
    \vspace{-7mm}
    \captionof{figure}{Failure case distribution across evolving times on Website and Game.}
    \label{fig:failurecase}
\end{minipage}

\noindent\textbf{EvoMAC continuously improves with the evolving times.} Fig.~\ref{fig:evolving} shows that as evolving iterations increase, performance consistently improves across all five dataset settings, covering two difficulty levels, two software types, and both requirement-oriented and function complement benchmarks. This highlights the effectiveness, generalizability, and robustness of the self-evolving approach, encouraging EvoMAC to evolve whenever possible.

\noindent\textbf{EvoMAC indistinguishably improves with different driving LLM.} From Fig.~\ref{fig:evolving}, we see that: i) both EvoMAC variants continuously improve with evolving iterations, demonstrating the robustness of the self-evolving design; ii) the two curves do not intersect, indicating that the EvoMAC variant powered by a more powerful single model consistently outperforms the other, highlighting the advantage of using a stronger model. Success builds on success.



\noindent\textbf{Failure case analysis.} 
Fig.~\ref{fig:failurecase} shows the failure case statistics across iterations for Website and Game, showing a general decrease in errors as iterations progress. We see that: i) the most common errors are page display issues in Website and logic errors in Game; ii) page errors are resolved more quickly, while logic errors persist, suggesting that more isolated issues are easier to fix during the evolution process. This results in a sharp initial performance improvement as sipler problems are addressed early, followed by a plateau as more complex issues remain unresolved, shown in Fig.~\ref{fig:evolving}.



\vspace{-4mm}
\subsection{Ablation study}
\vspace{-2mm}

To assess the effectiveness of each component, Tab.~\ref{Table:ablation study} details an ablation study featuring seven EvoMAC variants. 



\noindent\textbf{Effectiveness of objective environment feedback.}
Environment feedback, such as code execution logs, is essential for software development. Variant f) omits this tool, instead using an LLM-driven agent to critique the code. Comparing Variant g) with Variant f) shows a notable performance drop: Website tasks decrease by 12.67\% and 14.97\%, and Game tasks by 21.74\% and 18.28\% for Basic and Advanced levels, respectively. This underscores the importance of objective environmental feedback, as agent-driven critiques may introduce bias and fail to guide the evolution effectively.

\noindent\textbf{Effectiveness of multi-agent collaboration in coding team and testing team.} Comparing Variant g) to Variant e), we observe a performance decrease of 7.19\% and 5.26\% on Website Basic and Advanced respectively, when the coding team is reduced to a single agent. Similarly, comparing Variant g) to Variant d), there is a performance drop of 6.85\% and 6.48\% on Website Basic and Advanced respectively, also when the team is reduced to a single agent. These results demonstrate the necessary for involving multi-agent collaboration, highlighting that multi-agent setups offer more flexible adjustments and enhanced capabilities for evolution.




\begin{figure}[!t]
    \centering
    \includegraphics[width=0.95\linewidth]{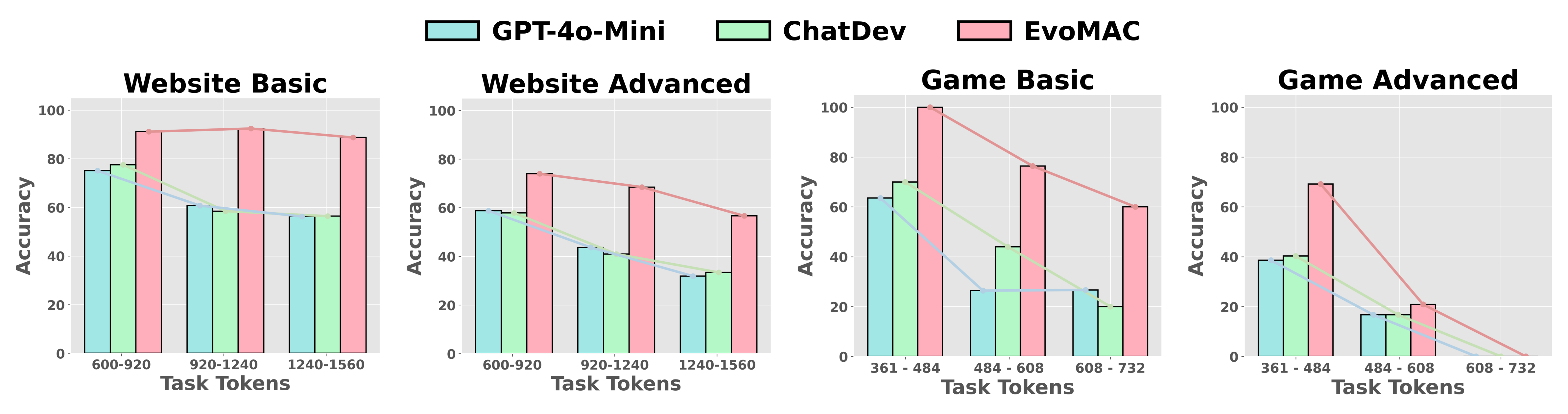}
    \vspace{-4mm}
    \caption{EvoMAC outperforms previous multi-agent and single-agent systems across all the context lengths across the four dataset settings on rSDE-Bench.}
    \vspace{-2mm}
    \label{fig:acc_dist}
\end{figure}



\noindent\textbf{Effectiveness in handling varied task token lengths.} Fig.~\ref{fig:acc_dist} shows a comparison of task token lengths and performance across GPT-4o-Mini, ChatDev, and EvoMAC. We see that: i) EvoMAC consistently outperforms ChatDev and GPT-4o-Mini across all context lengths, with its self-evolving mechanism enabling the identification and correction of missed contexts and errors during iterations; ii) EvoMAC experiences less performance degradation on the rSDE-Bench Website than on the Game, as Website tasks are more modular and can be broken into subtasks, whereas Game tasks require more coordinated management, making them more challenging.

\begin{figure}[!t]
    \centering
    \includegraphics[width=1.0\linewidth]{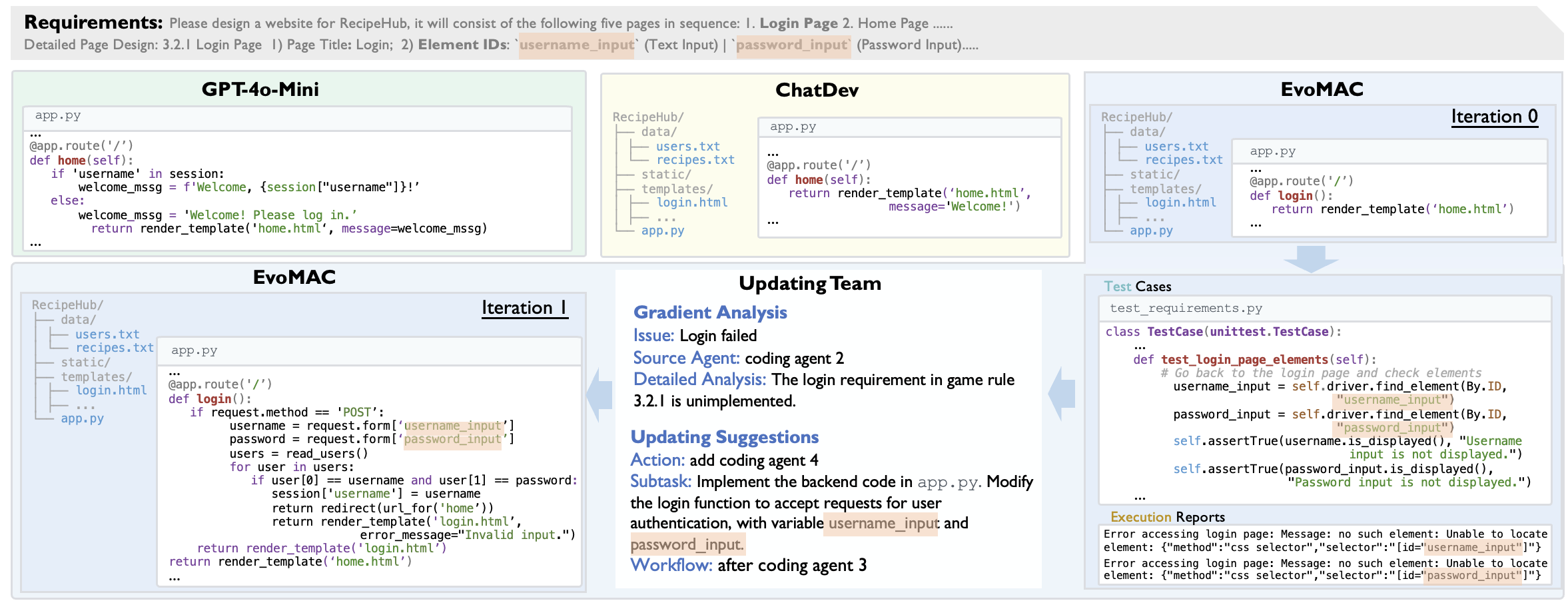}
    \vspace{-6mm}
    \caption{We show the generated code of single-agent, GPT-4o-Mini, and multi-agent systems, ChatDev, and our EvoMAC (iteration =0/1) given the Website task (RecipeHub). After evolving, EvoMAC can revise previous issues and fulfill the task requirement.}
    \vspace{-6mm}
    \label{fig:casestudy}
\end{figure}

\vspace{-4mm}
\subsection{Case study}
\vspace{-2mm}

Fig.~\ref{fig:casestudy} presents the generated code by a single agent, GPT-4o-Mini, multi-agent systems, ChatDev, and our EvoMAC before and after evolving (iteration=0/1). We see that: i) EvoMAC after evolving can correct issues from previous iterations and successfully fulfill the task requirements; ii) multi-agent systems tend to better comprehend the task requirements and produce more well-structured code. More generated software can refer to Sec.~\ref{app:software} in the Appendix.

\vspace{-4mm}
\section{Conclusion}
\vspace{-3mm}

We propose EvoMAC, a novel self-evolving paradigm for MAC networks. EvoMAC iteratively adapts agents and their connections during the testing phase of each task. It achieves this with a novel textual back-propagation algorithm. EvoMAC can push coding capabilities beyond function-level tasks and into more complex, software-level development. Furthermore, we propose rSDE-Bench, a novel requirement-oriented software development benchmark. rSDE-Bench features both complex and diverse software requirements, as well as the automatic evaluation of requirement correctness. Comprehensive experiments validate that the automatic requirement-aware evaluation in rSDE-Bench aligns closely with human evaluation. EvoMAC outperforms previous SOTAs in both software-level rSDE-Bench and function-level HumanEval benchmarks. 

\noindent\textbf{Future works.} In the future, we plan to introduce a reward model to enhance the self-evolving paradigm's ability to learn from feedback and extend the rSDE-Bench to more software types. 





{\small
\bibliographystyle{plainnat}
\bibliography{ref}

\begin{thebibliography}{35}
\providecommand{\natexlab}[1]{#1}
\providecommand{\url}[1]{\texttt{#1}}
\expandafter\ifx\csname urlstyle\endcsname\relax
  \providecommand{\doi}[1]{doi: #1}\else
  \providecommand{\doi}{doi: \begingroup \urlstyle{rm}\Url}\fi

\bibitem[Amazon(2022)]{CodeWhisperer}
Amazon.
\newblock {CodeWhisperer}.
\newblock In \emph{https://platform.qa.com/course/amazon-codewhisperer-generating-code-ai-4679/introduction}, 2022.
\newblock URL \url{https://platform.qa.com/course/amazon-codewhisperer-generating-code-ai-4679/introduction}.

\bibitem[Asai et~al.(2023)Asai, Wu, Wang, Sil, and Hajishirzi]{Asai2023SelfRAGLT}
Akari Asai, Zeqiu Wu, Yizhong Wang, Avirup Sil, and Hannaneh Hajishirzi.
\newblock Self-rag: Learning to retrieve, generate, and critique through self-reflection.
\newblock \emph{ArXiv}, abs/2310.11511, 2023.

\bibitem[Austin et~al.(2021)Austin, Odena, Nye, Bosma, Michalewski, Dohan, Jiang, Cai, Terry, Le, et~al.]{mbpp2021}
Jacob Austin, Augustus Odena, Maxwell Nye, Maarten Bosma, Henryk Michalewski, David Dohan, Ellen Jiang, Carrie Cai, Michael Terry, Quoc Le, et~al.
\newblock Program synthesis with large language models.
\newblock \emph{arXiv preprint arXiv:2108.07732}, 2021.

\bibitem[Chan et~al.(2024)Chan, Chen, Su, Yu, Xue, Zhang, Fu, and Liu]{chanchateval}
Chi-Min Chan, Weize Chen, Yusheng Su, Jianxuan Yu, Wei Xue, Shanghang Zhang, Jie Fu, and Zhiyuan Liu.
\newblock Chateval: Towards better llm-based evaluators through multi-agent debate.
\newblock In \emph{The Twelfth International Conference on Learning Representations}, 2024.

\bibitem[Chen et~al.(2021)Chen, Tworek, Jun, Yuan, de~Oliveira~Pinto, Kaplan, Edwards, Burda, Joseph, Brockman, Ray, Puri, Krueger, Petrov, Khlaaf, Sastry, Mishkin, Chan, Gray, Ryder, Pavlov, Power, Kaiser, Bavarian, Winter, Tillet, Such, Cummings, Plappert, Chantzis, Barnes, Herbert-Voss, Guss, Nichol, Paino, Tezak, Tang, Babuschkin, Balaji, Jain, Saunders, Hesse, Carr, Leike, Achiam, Misra, Morikawa, Radford, Knight, Brundage, Murati, Mayer, Welinder, McGrew, Amodei, McCandlish, Sutskever, and Zaremba]{humaneval2021}
Mark Chen, Jerry Tworek, Heewoo Jun, Qiming Yuan, Henrique~Ponde de~Oliveira~Pinto, Jared Kaplan, Harri Edwards, Yuri Burda, Nicholas Joseph, Greg Brockman, Alex Ray, Raul Puri, Gretchen Krueger, Michael Petrov, Heidy Khlaaf, Girish Sastry, Pamela Mishkin, Brooke Chan, Scott Gray, Nick Ryder, Mikhail Pavlov, Alethea Power, Lukasz Kaiser, Mohammad Bavarian, Clemens Winter, Philippe Tillet, Felipe~Petroski Such, Dave Cummings, Matthias Plappert, Fotios Chantzis, Elizabeth Barnes, Ariel Herbert-Voss, William~Hebgen Guss, Alex Nichol, Alex Paino, Nikolas Tezak, Jie Tang, Igor Babuschkin, Suchir Balaji, Shantanu Jain, William Saunders, Christopher Hesse, Andrew~N. Carr, Jan Leike, Josh Achiam, Vedant Misra, Evan Morikawa, Alec Radford, Matthew Knight, Miles Brundage, Mira Murati, Katie Mayer, Peter Welinder, Bob McGrew, Dario Amodei, Sam McCandlish, Ilya Sutskever, and Wojciech Zaremba.
\newblock Evaluating large language models trained on code.
\newblock 2021.

\bibitem[Chen et~al.(2023)Chen, Su, Zuo, Yang, Yuan, Qian, Chan, Qin, Lu, Xie, et~al.]{agentverse2023}
Weize Chen, Yusheng Su, Jingwei Zuo, Cheng Yang, Chenfei Yuan, Chen Qian, Chi-Min Chan, Yujia Qin, Yaxi Lu, Ruobing Xie, et~al.
\newblock Agentverse: Facilitating multi-agent collaboration and exploring emergent behaviors in agents.
\newblock \emph{arXiv preprint arXiv:2308.10848}, 2\penalty0 (4):\penalty0 6, 2023.

\bibitem[Google(2023)]{Codey}
Google.
\newblock {Codey}.
\newblock In \emph{https://console.cloud.google.com/vertex-ai/publishers/google/model-garden/codechat-bison}, 2023.
\newblock URL \url{https://console.cloud.google.com/vertex-ai/publishers/google/model-garden/codechat-bison}.

\bibitem[Hong et~al.(2023)Hong, Zheng, Chen, Cheng, Wang, Zhang, Wang, Yau, Lin, Zhou, et~al.]{metagpt2023}
Sirui Hong, Xiawu Zheng, Jonathan Chen, Yuheng Cheng, Jinlin Wang, Ceyao Zhang, Zili Wang, Steven Ka~Shing Yau, Zijuan Lin, Liyang Zhou, et~al.
\newblock Metagpt: Meta programming for multi-agent collaborative framework.
\newblock \emph{arXiv preprint arXiv:2308.00352}, 2023.

\bibitem[Hua et~al.(2023)Hua, Fan, Li, Mei, Ji, Ge, Hemphill, and Zhang]{waragent2023}
Wenyue Hua, Lizhou Fan, Lingyao Li, Kai Mei, Jianchao Ji, Yingqiang Ge, Libby Hemphill, and Yongfeng Zhang.
\newblock War and peace (waragent): Large language model-based multi-agent simulation of world wars.
\newblock \emph{arXiv preprint arXiv:2311.17227}, 2023.

\bibitem[Islam et~al.(2024)Islam, Ali, and Parvez]{mapcoder2024}
Md.~Ashraful Islam, Mohammed~Eunus Ali, and Md~Rizwan Parvez.
\newblock Mapcoder: Multi-agent code generation for competitive problem solving, 2024.
\newblock URL \url{https://arxiv.org/abs/2405.11403}.

\bibitem[Jimenez et~al.(2023)Jimenez, Yang, Wettig, Yao, Pei, Press, and Narasimhan]{swebench2023}
Carlos~E Jimenez, John Yang, Alexander Wettig, Shunyu Yao, Kexin Pei, Ofir Press, and Karthik~R Narasimhan.
\newblock Swe-bench: Can language models resolve real-world github issues?
\newblock In \emph{The Twelfth International Conference on Learning Representations}, 2023.

\bibitem[Khan et~al.(2023)Khan, Bari, Do, Wang, Parvez, and Joty]{xcodeeval}
Mohammad Abdullah~Matin Khan, M~Saiful Bari, Xuan~Long Do, Weishi Wang, Md~Rizwan Parvez, and Shafiq Joty.
\newblock xcodeeval: A large scale multilingual multitask benchmark for code understanding, generation, translation and retrieval, 2023.
\newblock URL \url{https://arxiv.org/abs/2303.03004}.

\bibitem[Li et~al.(2023)Li, Hammoud, Itani, Khizbullin, and Ghanem]{li2023camel}
Guohao Li, Hasan Abed Al~Kader Hammoud, Hani Itani, Dmitrii Khizbullin, and Bernard Ghanem.
\newblock Camel: Communicative agents for "mind" exploration of large language model society.
\newblock In \emph{Thirty-seventh Conference on Neural Information Processing Systems}, 2023.

\bibitem[Li et~al.(2024{\natexlab{a}})Li, Wang, Zheng, and Zhang]{li2024looglelongcontextlanguagemodels}
Jiaqi Li, Mengmeng Wang, Zilong Zheng, and Muhan Zhang.
\newblock Loogle: Can long-context language models understand long contexts?, 2024{\natexlab{a}}.
\newblock URL \url{https://arxiv.org/abs/2311.04939}.

\bibitem[Li et~al.(2024{\natexlab{b}})Li, Wang, Zhang, Li, Lai, Kang, Ma, and Liu]{agenthospital}
Junkai Li, Siyu Wang, Meng Zhang, Weitao Li, Yunghwei Lai, Xinhui Kang, Weizhi Ma, and Yang Liu.
\newblock Agent hospital: A simulacrum of hospital with evolvable medical agents, 2024{\natexlab{b}}.
\newblock URL \url{https://arxiv.org/abs/2405.02957}.

\bibitem[Li et~al.(2022{\natexlab{a}})Li, Choi, Chung, Kushman, Schrittwieser, Leblond, Tom, Eccles, Keeling, Gimeno, Lago, Hubert, Choy, de, d’Autume, Babuschkin, Chen, Huang, Welbl, Gowal, Alexey, Cherepanov, Molloy, Mankowitz, Robson, Kohli, de, Freitas, Kavukcuoglu, and Vinyals]{Li2022CompetitionlevelCG}
Yujia Li, David Choi, Junyoung Chung, Nate Kushman, Julian Schrittwieser, R{\'e}mi Leblond, Tom, Eccles, James Keeling, Felix Gimeno, Agustin~Dal Lago, Thomas Hubert, Peter Choy, Cyprien de, Masson d’Autume, Igor Babuschkin, Xinyun Chen, Po-Sen Huang, Johannes Welbl, Sven Gowal, Alexey, Cherepanov, James Molloy, Daniel~Jaymin Mankowitz, Esme~Sutherland Robson, Pushmeet Kohli, Nando de, Freitas, Koray Kavukcuoglu, and Oriol Vinyals.
\newblock Competition-level code generation with alphacode.
\newblock \emph{Science}, 378:\penalty0 1092 -- 1097, 2022{\natexlab{a}}.

\bibitem[Li et~al.(2022{\natexlab{b}})Li, Choi, Chung, Kushman, Schrittwieser, Leblond, Eccles, Keeling, Gimeno, Dal~Lago, Hubert, Choy, de~Masson~d’Autume, Babuschkin, Chen, Huang, Welbl, Gowal, Cherepanov, Molloy, Mankowitz, Sutherland~Robson, Kohli, de~Freitas, Kavukcuoglu, and Vinyals]{Li_2022}
Yujia Li, David Choi, Junyoung Chung, Nate Kushman, Julian Schrittwieser, Rémi Leblond, Tom Eccles, James Keeling, Felix Gimeno, Agustin Dal~Lago, Thomas Hubert, Peter Choy, Cyprien de~Masson~d’Autume, Igor Babuschkin, Xinyun Chen, Po-Sen Huang, Johannes Welbl, Sven Gowal, Alexey Cherepanov, James Molloy, Daniel~J. Mankowitz, Esme Sutherland~Robson, Pushmeet Kohli, Nando de~Freitas, Koray Kavukcuoglu, and Oriol Vinyals.
\newblock Competition-level code generation with alphacode.
\newblock \emph{Science}, 378\penalty0 (6624):\penalty0 1092–1097, December 2022{\natexlab{b}}.
\newblock ISSN 1095-9203.
\newblock \doi{10.1126/science.abq1158}.
\newblock URL \url{http://dx.doi.org/10.1126/science.abq1158}.

\bibitem[Liu et~al.(2023)Liu, Xia, Wang, and ZHANG]{evalplus}
Jiawei Liu, Chunqiu~Steven Xia, Yuyao Wang, and LINGMING ZHANG.
\newblock Is your code generated by chat{GPT} really correct? rigorous evaluation of large language models for code generation.
\newblock In \emph{Thirty-seventh Conference on Neural Information Processing Systems}, 2023.
\newblock URL \url{https://openreview.net/forum?id=1qvx610Cu7}.

\bibitem[Mandi et~al.(2024{\natexlab{a}})Mandi, Jain, and Song]{mandi2024roco}
Zhao Mandi, Shreeya Jain, and Shuran Song.
\newblock Roco: Dialectic multi-robot collaboration with large language models.
\newblock In \emph{2024 IEEE International Conference on Robotics and Automation (ICRA)}, pages 286--299. IEEE, 2024{\natexlab{a}}.

\bibitem[Mandi et~al.(2024{\natexlab{b}})Mandi, Jain, and Song]{roco2024}
Zhao Mandi, Shreeya Jain, and Shuran Song.
\newblock Roco: Dialectic multi-robot collaboration with large language models.
\newblock In \emph{2024 IEEE International Conference on Robotics and Automation (ICRA)}, pages 286--299. IEEE, 2024{\natexlab{b}}.

\bibitem[Microsoft(2023)]{Copilot}
Microsoft.
\newblock {Copilot}.
\newblock In \emph{https://www.microsoft.com/en-us/microsoft-copilot/meet-copilot}, 2023.
\newblock URL \url{https://www.microsoft.com/en-us/microsoft-copilot/meet-copilot}.

\bibitem[Osika(2023)]{GPTEngineer}
Anton Osika.
\newblock {GPT-Engineer}.
\newblock In \emph{https://github.com/AntonOsika/gpt-engineer}, 2023.
\newblock URL \url{https://github.com/AntonOsika/gpt-engineer}.

\bibitem[Pang et~al.(2024)Pang, Tang, Ye, Xiong, Zhang, Wang, and Chen]{matrix2024}
Xianghe Pang, Shuo Tang, Rui Ye, Yuxin Xiong, Bolun Zhang, Yanfeng Wang, and Siheng Chen.
\newblock Self-alignment of large language models via monopolylogue-based social scene simulation.
\newblock In \emph{Forty-first International Conference on Machine Learning}, 2024.

\bibitem[Qian et~al.(2023)Qian, Liu, Liu, Chen, Dang, Li, Yang, Chen, Su, Cong, Xu, Li, Liu, and Sun]{chatdev2023}
Chen Qian, Wei Liu, Hongzhang Liu, Nuo Chen, Yufan Dang, Jiahao Li, Cheng Yang, Weize Chen, Yusheng Su, Xin Cong, Juyuan Xu, Dahai Li, Zhiyuan Liu, and Maosong Sun.
\newblock Chatdev: Communicative agents for software development.
\newblock \emph{arXiv preprint arXiv:2307.07924}, 2023.
\newblock URL \url{https://arxiv.org/abs/2307.07924}.

\bibitem[Valmeekam et~al.(2023)Valmeekam, Marquez, and Kambhampati]{Valmeekam2023CanLL}
Karthik Valmeekam, Matthew Marquez, and Subbarao Kambhampati.
\newblock Can large language models really improve by self-critiquing their own plans?
\newblock \emph{ArXiv}, abs/2310.08118, 2023.

\bibitem[Wang et~al.(2024{\natexlab{a}})Wang, Duan, Zhang, Lin, and Chen]{wang2024adaleval}
Chonghua Wang, Haodong Duan, Songyang Zhang, Dahua Lin, and Kai Chen.
\newblock Ada-leval: Evaluating long-context llms with length-adaptable benchmarks, 2024{\natexlab{a}}.

\bibitem[Wang et~al.(2024{\natexlab{b}})Wang, Salmani, Omidi, Ren, Rezagholizadeh, and Eshaghi]{wang2024limitssurveytechniquesextend}
Xindi Wang, Mahsa Salmani, Parsa Omidi, Xiangyu Ren, Mehdi Rezagholizadeh, and Armaghan Eshaghi.
\newblock Beyond the limits: A survey of techniques to extend the context length in large language models, 2024{\natexlab{b}}.
\newblock URL \url{https://arxiv.org/abs/2402.02244}.

\bibitem[Wu et~al.(2023)Wu, Bansal, Zhang, Wu, Li, Zhu, Jiang, Zhang, Zhang, Liu, Awadallah, White, Burger, and Wang]{autogen2023}
Qingyun Wu, Gagan Bansal, Jieyu Zhang, Yiran Wu, Beibin Li, Erkang Zhu, Li~Jiang, Xiaoyun Zhang, Shaokun Zhang, Jiale Liu, Ahmed~Hassan Awadallah, Ryen~W White, Doug Burger, and Chi Wang.
\newblock Autogen: Enabling next-gen llm applications via multi-agent conversation, 2023.
\newblock URL \url{https://arxiv.org/abs/2308.08155}.

\bibitem[Xu et~al.(2024)Xu, Liu, Liu, Hou, Li, Zhang, Wang, Zeng, Du, Zhao, Tang, and Dong]{Xu2024ChatGLMMathIM}
Yifan Xu, Xiao Liu, Xinghan Liu, Zhenyu Hou, Yueyan Li, Xiaohan Zhang, Zihan Wang, Aohan Zeng, Zhengxiao Du, Wenyi Zhao, Jie Tang, and Yuxiao Dong.
\newblock Chatglm-math: Improving math problem-solving in large language models with a self-critique pipeline.
\newblock \emph{ArXiv}, 2024.

\bibitem[Xu et~al.(2023)Xu, Wang, Li, Luo, Wang, Liu, and Liu]{werewolf2023}
Yuzhuang Xu, Shuo Wang, Peng Li, Fuwen Luo, Xiaolong Wang, Weidong Liu, and Yang Liu.
\newblock Exploring large language models for communication games: An empirical study on werewolf.
\newblock \emph{arXiv preprint arXiv:2309.04658}, 2023.

\bibitem[Yang et~al.(2024{\natexlab{a}})Yang, Zhou, Chen, and Zhang]{CodeScore}
Guang Yang, Yu~Zhou, Xiang Chen, and Xiangyu Zhang.
\newblock Codescore-r: An automated robustness metric for assessing the functionalcorrectness of code synthesis, 2024{\natexlab{a}}.
\newblock URL \url{https://arxiv.org/abs/2406.06902}.

\bibitem[Yang et~al.(2024{\natexlab{b}})Yang, Jimenez, Wettig, Lieret, Yao, Narasimhan, and Press]{Yang2024SWEagentAI}
John Yang, Carlos~E. Jimenez, Alexander Wettig, Kilian Lieret, Shunyu Yao, Karthik Narasimhan, and Ofir Press.
\newblock Swe-agent: Agent-computer interfaces enable automated software engineering.
\newblock \emph{ArXiv}, abs/2405.15793, 2024{\natexlab{b}}.

\bibitem[Zheng et~al.(2023)Zheng, Ning, Chen, Wang, Chen, Guo, and Wang]{Zheng2023TowardsAU}
Zibin Zheng, Kai-Chun Ning, Jiachi Chen, Yanlin Wang, Wenqing Chen, Lianghong Guo, and Weicheng Wang.
\newblock Towards an understanding of large language models in software engineering tasks.
\newblock \emph{ArXiv}, abs/2308.11396, 2023.

\bibitem[Zhou et~al.(2024)Zhou, Ou, Ding, Li, Wu, Wang, Chen, Wang, Xu, Zhang, et~al.]{symbolic2024}
Wangchunshu Zhou, Yixin Ou, Shengwei Ding, Long Li, Jialong Wu, Tiannan Wang, Jiamin Chen, Shuai Wang, Xiaohua Xu, Ningyu Zhang, et~al.
\newblock Symbolic learning enables self-evolving agents.
\newblock \emph{arXiv preprint arXiv:2406.18532}, 2024.

\bibitem[Ziems et~al.(2024)Ziems, Held, Shaikh, Chen, Zhang, and Yang]{ziems2024can}
Caleb Ziems, William Held, Omar Shaikh, Jiaao Chen, Zhehao Zhang, and Diyi Yang.
\newblock Can large language models transform computational social science?
\newblock \emph{Computational Linguistics}, 50\penalty0 (1):\penalty0 237--291, 2024.

\end{thebibliography}
}

\newpage
\section*{Appendix}

\section{Benchmark details}
\label{app:benchmark}

\begin{minipage}{0.5\textwidth}
\centering
\captionof{table}{Basic statistics for website and game domains, including the amount of samples, length in lines of code (Basic/Advanced), and number of test cases at both Basic and Advanced levels.}
\scalebox{0.8}{
\begin{tabular}{lcccc}
\toprule
\multirow{2}{*}{Benchmark} & \multicolumn{2}{c}{Software} & \multicolumn{2}{c}{Test Case} \\
                           & Amount   & Length  & Basic        & Advanced        \\ \midrule
Website                    & 45       &         1011/1553           &        292      &         247        \\ 
Game                       & 8        &         507/788           &        46      &         31        \\
\bottomrule
\end{tabular}}
\label{Table:benchmark_stat}
\end{minipage}
\begin{minipage}{0.5\textwidth}
    \centering
    \includegraphics[width=1.0\linewidth]{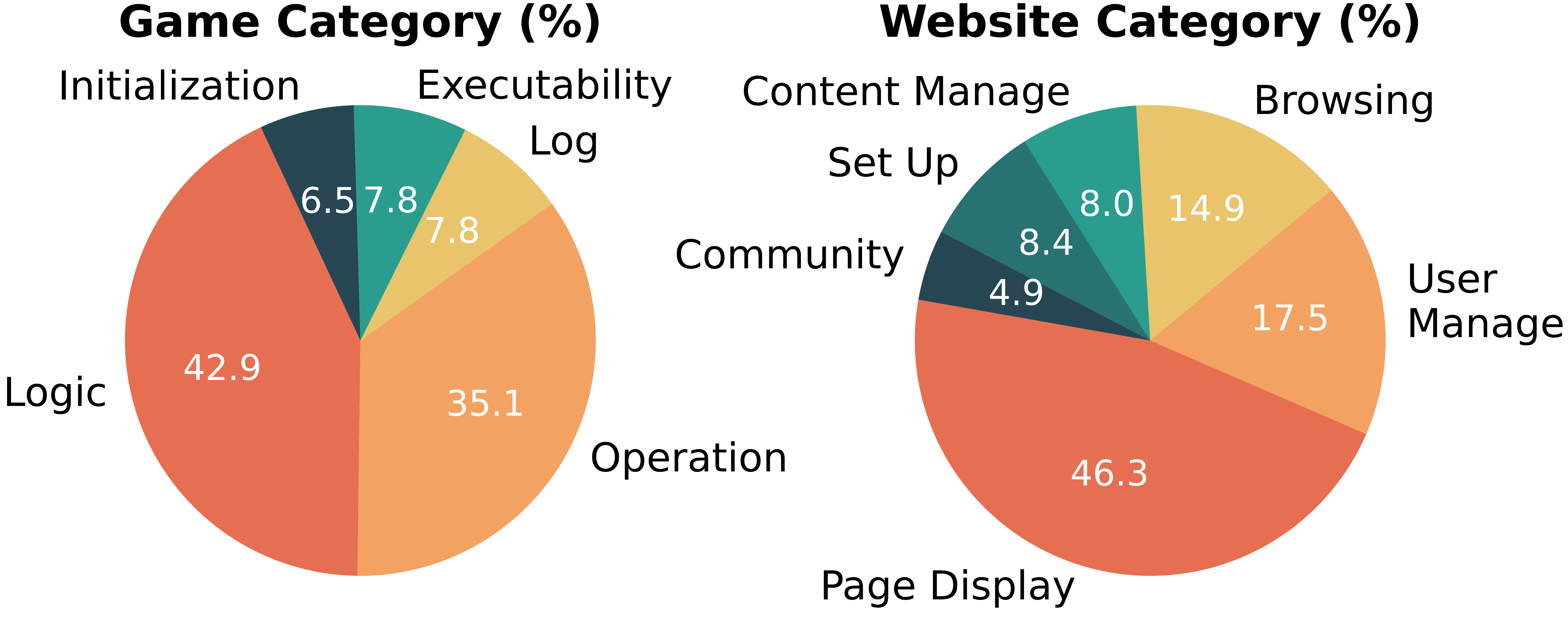}
    \vspace{-7mm}
    \captionof{figure}{Statistics of Game and Website tasks.}
    \label{Fig:testcase_dist}
\end{minipage}

\noindent\textbf{Step 1: Software requirement generation.} 
Each task instance begins with the generation of clear, measurable software requirements. Given the inherent differences across various types of software, we adopt distinct approaches for their formulation. For game-related software, we focuses on common real-world games, capturing detailed task requirements such as GUI layout initialization, interaction methods, and game rules. To align more closely with actual game development practices, we also include game state logging as part of the software requirements. Due to the complexity of logic in game software, these requirements are manually crafted by human. In contrast, for website-related software, we begin with a concise website name, and then leverage the large language model (\texttt{gpt-4o-mini}) to enrich the requirements according to predefined patterns. This approach ensures both efficiency and scalability in the creation of benchmarks for websites. By tailoring the process to the distinct characteristics of each software domain, we maintain precision in requirement formulation while addressing the unique challenges posed by each context.

\noindent\textbf{Step 2: Requirement-based test cases generation.} 

\begin{figure}[!t]
    \centering
    \includegraphics[width=1.0\linewidth]{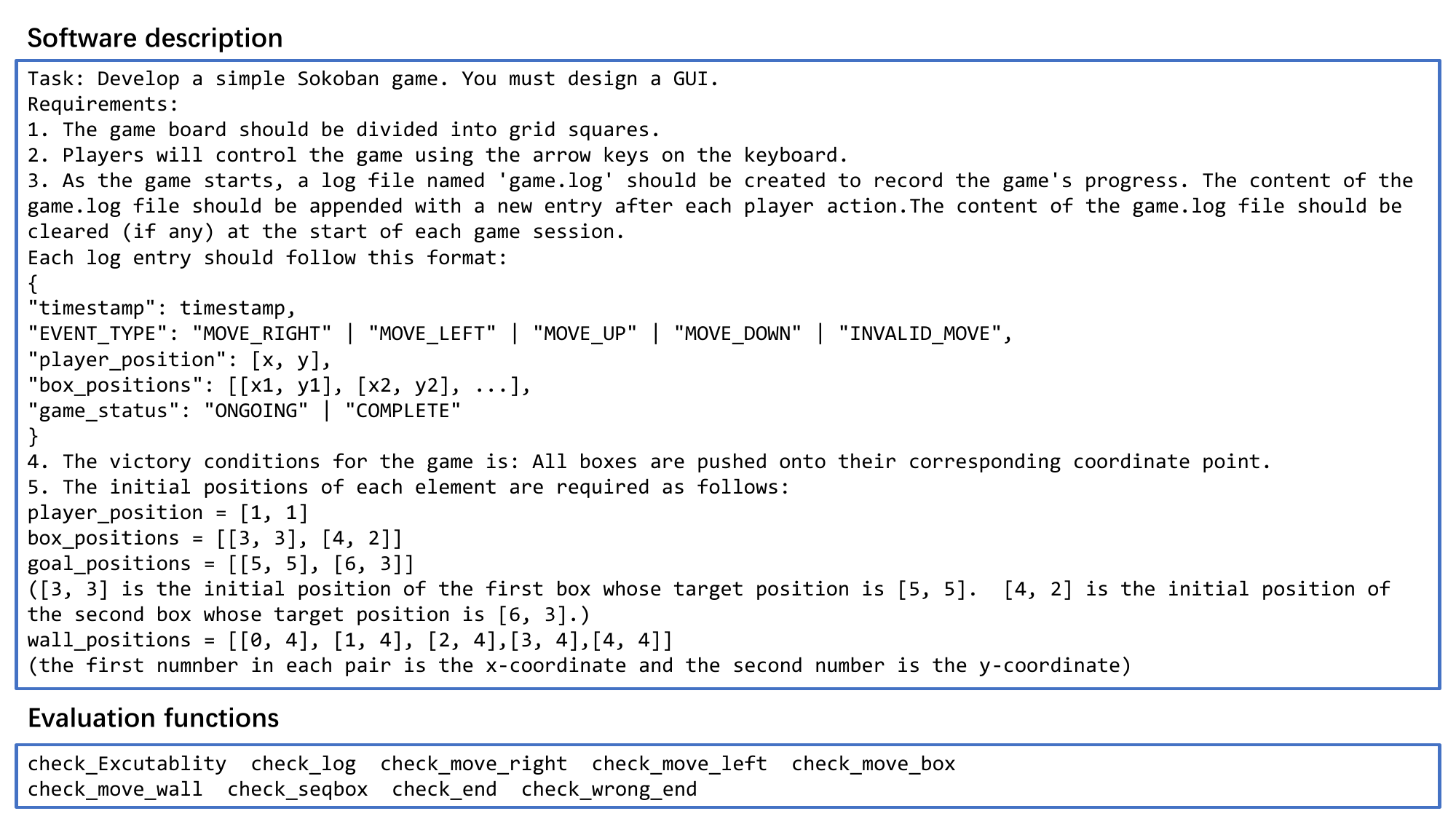}
    \caption{Test cases of Game in rSDE-Bench.}
    \label{fig:game_requirement}
\end{figure}

\begin{figure}[!t]
    \centering
    \includegraphics[width=1.0\linewidth]{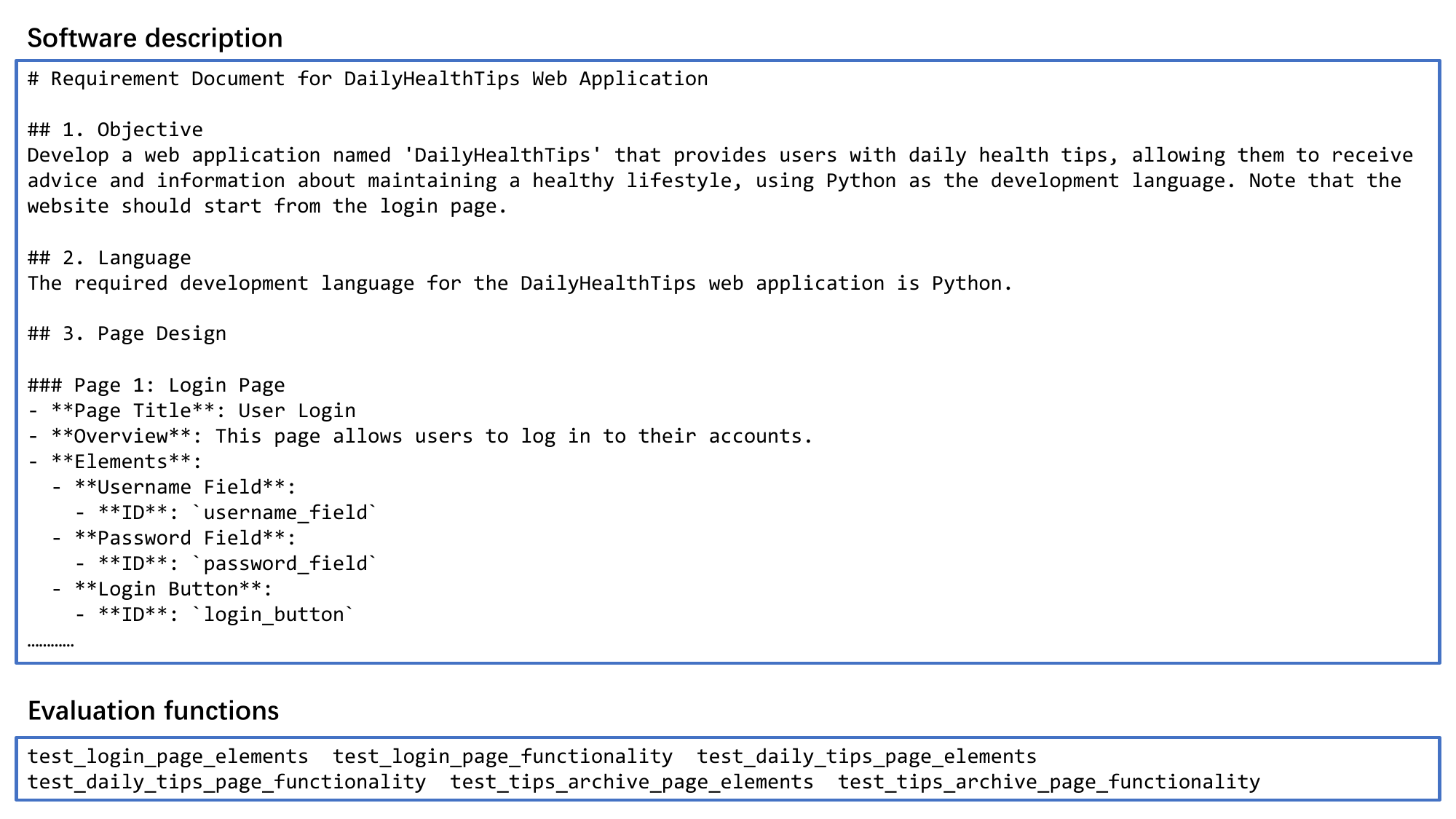}
    \caption{Test cases of Website in rSDE-Bench.}
    \label{fig:web_requirement}
\end{figure}

As illustrated in Fig~\ref{fig:game_requirement} and Fig~\ref{fig:web_requirement}, unit tests offer a precise evaluation of software completion. Each task instance includes black-box unit test cases that correspond directly to the software requirements, allowing for a quantitative assessment of requirement fulfillment. To further assess the model's code generation capabilities, we categorize test cases into two levels of difficulty—basic and advanced, as outlined in Tab.~\ref{Table:benchmark_stat}. We also provide an overview of all websites and games in Tab.~\ref{tab:all websites} and Tab.~\ref{tab:all games} respectively. As shown in Fig.~\ref{Fig:testcase_dist}, test cases for website and game software exhibit structural differences, reflecting the distinct nature of each software type. They enable more targeted evaluation of code generation capabilities. Thus, similar to software requirements, the test cases are constructed differently based on the software type. For game-related tests, we manually create test cases, akin to the HumanEval~\cite{humaneval2021} benchmark, which tracks state changes in response to specific inputs. In the game environment, we assess how game states evolve in response to GUI interactions. For website-related tests, large language model (\texttt{gpt-4o-mini}) generates Selenium-based test cases aligned with the software requirements, followed by manual corrections to resolve any ambiguities. This structured approach ensures rigorous evaluation across diverse software domains.

\textbf{Basic and advanced requirements definition.} For the games, basic requirements involve straightforward user interactions that do not require complex logic, such as character movement or interacting with simple GUI elements. Advanced requirements incorporate more intricate logic, such as managing game state transitions based on user actions or handling conditional game events. These cases focus on ensuring the correct execution of basic actions. In contrast, advanced cases incorporate more intricate logic, such as managing game state transitions based on user actions or handling conditional game events. These cases challenge the model's ability to generate code that integrates dynamic decision-making and interaction within the game environment. For websites, basic cases focus on ensuring that the necessary page elements—such as input fields, buttons, and layouts—are present correctly. These cases assess the completeness of the webpage's structure. On the other hand, advanced cases evaluate more complex functionality, such as handling user authentication, managing dynamic content, or executing specific operations within a content management system. These cases require the model to generate code that performs backend logic and manages user interactions at a deeper level.

\begin{table}[]
\centering
\caption{Overview of Websites in rSDE-Bench.}
\begin{tabular}{lll}
\hline
\multicolumn{3}{c}{\textbf{Websites}} \\ \hline
CharitableGivingPlatform      & DailyHealthTips             & DailyJournalApp         \\
EcoFriendlyLivingTips         & ElderCareResources          & EventPlanner            \\
FitnessEquipmentRental        & FitnessTracker              & FreelancerMarketplace   \\
GreenLivingGuide              & HealthConsultationPlatform  & MotivationalQuotesApp   \\
MusicFestivalDirectory        & NoteTakingApp               & NutritionInformationHub \\
OnlineLibraryManagementSystem & OnlineTherapeuticJournaling & OnlineThriftStore       \\
PeerTutoringNetwork           & PersonalBlog                & PersonalFinanceBlog     \\
RecipeHub                     & RemoteInternshipMarketplace & RemoteJobBoard          \\
TravelDiary                   & VirtualBookPublishing       & VirtualWellnessRetreats \\
DigitalArtworkGallery         & DigitalStorytellingPlatform & ExpenseTracker          \\
FitnessChallenges             & GardeningForBeginners       & GourmetFoodSubscription \\
MovieRecommendationSystem     & MusicCollaborator           & OnlineCulturalExchange  \\
OnlineCulturalFestivals       & OnlineVintageMarket         & ParentingAdviceForum    \\
PetCareCommunity              & PortfolioSite               & SkillShare              \\
TaskManager                   & VolunteerMatch              & OnlineShoppingCenter   \\ \hline
\end{tabular}
\label{tab:all websites}
\end{table}
\begin{table}[]
\centering
\caption{Overview of Games in rSDE-Bench.}
\begin{tabular}{llll}
\hline
\multicolumn{4}{c}{\textbf{Games}} \\ \hline
Balls & Tank      & Racing  & Ghostly \\
Mario & Bomberman & Sokoban & Brick  \\ \hline
\end{tabular}
\label{tab:all games}
\end{table}

\section{Algorithm}
In this section, we present the algorithm of EvoMAC in Alg.~\ref{alg:EvoMAC}. For more details, please refer to Section 3.

\begin{algorithm}[!t]
	\caption{Self-Evolving Paradigm} 
	\begin{algorithmic}[1]
        \Require $\mathbf{X}$ \Comment{Task input}
        \Require $\mathcal{A}_{g}^{(0)}$ \Comment{Initialized MAC network: agent prompts and pipeline}
        \Require $\mathcal{A}_{t}$ \Comment{Designed MAC network to generate target proxy}
        \Require $\mathcal{G}$ \Comment{Agent-based gradient function}
        \Require $\mathcal{U}$ \Comment{Agent-based update function}
        \Require $E$ \Comment{Environment tool to generate loss}
        \State Define $K$ as the number of self-evolving iterations, $\Phi$ as MACN generation process
        \State {\color{blue} \#~Target Proxy}
        \State $\mathbf{T} = \Phi(\mathbf{X},\mathcal{A}_{t})$ 
        \State {\color{blue} \#~Self-Evolving Procedure}
        \For {$k = 0, 1, \dots, K-1$}
            \State {\color{blue} \#~Forward Pass}
            \State $\mathbf{G}^{(k)} = \Phi(\mathbf{X}, \mathcal{A}_g^{(k)})$
            \State {\color{blue} \#~Loss Computation}
            \State $\mathbf{L}^{(k)}=\langle \mathbf{G}^{(k)}, \mathbf{T}\rangle_{E}$ \Comment{Use environment feedback as textual loss}
            \State {\color{blue} \#~Textual Backpropagation}
            \State $\nabla \mathbf{L}^{(k)} = \mathcal{G}(\mathbf{L}^{(k)},\mathcal{A}_g^{(k)})$ \Comment{Summarize textual gradient}
            \State $\mathcal{A}_g^{(k+1)} = \mathcal{U}(\mathcal{A}_g^{(k)}, \nabla \mathbf{L}^{(k)})$ \Comment{Update agent prompts and pipeline}
        \EndFor
        \State \Return $\mathcal{A}_g^{(K)}, \mathbf{G}^{(K)}$
	\end{algorithmic}
	\label{alg:EvoMAC}
\end{algorithm}

\section{Prompts}
\label{app:prompts}
In this section, we present the agents' prompts in EvoMAC, including coding organizer (Tab.~\ref{tab:CodingOrganizer}), coding agent (Tab.~\ref{tab:CodingAgent}), testing organizer (Tab.~\ref{tab:TestingOrganizer}), testing agent (Tab.~\ref{tab:TestingAgent}), gradient agent (Tab.~\ref{tab:GradientAgent}), and update agent (Tab.~\ref{tab:UpdatingAgent}).
  
\begin{table}[!ht]
\caption{\textbf{Coding Organizer}}
\label{tab:CodingOrganizer}
\begin{response}
\small
According to the new user's task and our software designs listed below: 

Task: "\{task\}".

Task description: "\{description\}".

Modality: "\{modality\}".

Programming Language: "\{language\}"

Requirements analysis: "\{requirements\}"

Ideas:"\{ideas\}"

Coding plan: "\{codes\}"

Your goal is to organize a coding team to complete the software development task.

There are two default tasks: 

1) log the user's actions and events in a game.log file according to the task requirements in the write format! Be careful and make sure to maintain the game.log file right! The log should happened after the action is taken, record the most recent state.

2) create a user interface (GUI) for the game using the programming language and the requirements analysis. The GUI should be beautiful and user-friendly.

Besides these tasks, you should pay attention to the unachieved requirements and think step by step to formulate the requirements into concrete tasks.

You should follow the following format: \"COMPOSITION\" is the composition of tasks, and \"Workflow\" is the workflow of the programmers. Each task is assigned to a programmer, and the workflow shows the dependencies between tasks. 

\#\#\# COMPOSITION

```

Task 1: Task 1 description

Task 2: Task 2 description

```

\#\#\# WORKFLOW

```

Task 1: []

Task 2: [Task 1]

```

Please note that the decomposition should be both effective and efficient.

1) Each decomposed task should include the related the functions. The task description should be clear and concise. 

2) The composition should be kept as small as possible! (LESS THAN "\{num\_agents\}"). If there are more than 5 tasks, consider merging the tasks and focus on the most essential features. 

3) The decomposed tasks should fully cover the task definitions.

4) The workflow should not contain circles!

\end{response}
\end{table}

\begin{table}[!ht]
\caption{\textbf{Coding Agent}}
\label{tab:CodingAgent}
\begin{response}
\small
According to the new user's task and our software designs listed below: 

Task: "\{task\}".

Modality: "\{modality\}".

Programming Language: "\{language\}"

Sub-Task description: "\{subtask\}"

Codes:

"\{codes\}"

Unimplemented File:

"\{unimplemented\_file\}"

Your first think step by step first reason yourself about the files and functions related to the sub-task.

Then you should output the COMPLETE code content in each file. Each file must strictly follow a markdown code block format, where the following tokens must be replaced such that "FILENAME" is the lowercase file name including the file extension, "LANGUAGE" in the programming language, "DOCSTRING" is a string literal specified in source code that is used to document a specific segment of code, and "CODE" is the original code. Format:

FILENAME

```LANGUAGE

'''

DOCSTRING

'''

CODE

```

Implementation Requirements:

1. As the {assistant\_role}, to satisfy the complete function of our developed software, you have to implement all functions which are related to the subtask.

2. If the function is implemented, recheck the logic and log to ensure the targeted feature is fully achieved

3. Important: that both the logic and log should be fully functional! No placeholder (such as 'pass' in Python), strictly following the required format. You must strictly following the required format. You must strictly following the required format.

4. Ensure the functions are consistent among different files, and correctely imported.

Additional Note: {additional\_note}

\end{response}
\end{table}

\begin{table}[!ht]
\caption{\textbf{Testing Organizer}}
\label{tab:TestingOrganizer}
\begin{response}
\small
According to the software requirements listed below: 

Task: "\{task\}".

Modality: "\{modality\}".

Programming Language: "\{language\}"

Your goal is to organize a testing team to complete the software development task.

There are four default tasks: 

1) carefully test the logging mechanism according to the task requirements! The log should happened immediately after the action is taken, record the most recent state. Remember the logging order is very important, record basic operation first then record the subsequent events. Ensure the data format, keys and values are accurate and right! Pay attention to the nested data type and carefully check each element.

2) test the logging mechanism for the special triggered conditions.

3) test the value initialziation required by the task are correctly achieved, pay attention to the corrdinates.

4) test the function inputs and the global variable are imported in each functions, ensure the input values and global variable used in the function are valid and involved when the function is called.

5) test each event in the task is implemented and that the logic triggered matches the conditions in the task description.

Follow the format: "COMPOSITION" is the composition of tasks, and "Workflow" is the workflow of the programmers. 

\#\#\# COMPOSITION

```

Task 1: Task 1 description

Task 2: Task 2 description

```

\#\#\# WORKFLOW

```

Task 1: []

Task 2: [Task 1]

```

\end{response}
\end{table}

\begin{table}[!ht]
\caption{\textbf{Testing Agent}}
\label{tab:TestingAgent}
\begin{response}
\small
Our software requirements and developed source codes are listed below: 

Programming Language: "\{language\}"

Source Codes:

"\{codes\}"

Testing Task description: "\{subtask\}"

According to Testing Task description, please write test cases to locate the bugs, note that logging is important, ensure the content and format is right. 

You should first locate the functions that need to be tested and write the test cases for them according to the testing task description.

The output must strictly follow a markdown code block format, where the following tokens must be replaced such that "FILENAME" is "{test\_file\_name}", "LANGUAGE" in the programming language,"REQUIREMENTS" is the targeted requirement of the test case, and "CODE" is the test code that is used to test the specific requirement of the file. Format:

FILENAME

```LANGUAGE

'''

REQUIREMENTS

'''

CODE

```

You will start with the "{test\_file\_name}" and finish the code follows in the strictly defined format.

Please note that:

1) The code should be fully functional. Ensure to implement all functions. No placeholders (such as 'pass' in Python).

2) You should not write anything about log testing unless testing task description clearly state that the logs need to be tested

3) You should write the test file with 'unittest' python library.

4) You should not modify the source code, only write the test code. Very Important!

\end{response}
\end{table}

\begin{table}[!ht]
\caption{\textbf{Gradient Agent}}
\label{tab:GradientAgent}
\begin{response}
\small
Our developed source codes and corresponding test reports are listed below: 

Programming Language: "\{language\}"

User requirement:

"\{task\}"

Source Codes:

"\{codes\}"

The execution outcome of our source codes:

"\{test\_reports\}"

We also have write test case to test our source codes, our test codes are listed below: 

"\{test\_codes\}"

And the execution outcome of our test codes is: 

"\{testcase\_reports\}"

According to these imformation, please analyze the source code, test code and execution reports. Make sure your analysis aligns with the source code and user requirements.

First, determine whether the error is caused by incorrect test code; if so, respond "Wrong test code." The wrong type of test code includes not matching the user requirement, e.g. wrong value of the test reference answer conflict with user requirement desciption or improper use of source code. Please be careful and not make wrong judgment about the source code.

Second, if there exist bugs in the source code, give a detailed analysis of the problem. Your answer MUST follow the format below:

file name:file\_1.py

function name: function\_1, function\_2

detailed analysis of the problem: your analysis

file name:file\_2.py

function name: function\_3, function\_4

detailed analysis of the problem: your analysis

Your answer should also follow the requirements below:

1) The answer should only include the analysis of the wrong source code. Please not include analysis of the testcase error here. If all the reports show that there is no bugs in source codes and test codes, you should just only reply: No error in codes.

2) You can answer more than one function name, but you can only answer one file name each time. If you want to answer two file names, you should split it and answer with the format respectively.(Answer the file\_1.py and corresponding information with the format, and then answer the file\_2.py and corresponding information with the format)

3) You may include one or more function names in each file, but you should not include the same function name in different files.

4) You should not answer anything about test file(e.g. file name: test\_requirement\_0.py) in your answer. VERY IMPORTANT!

\end{response}
\end{table}

\begin{table}[!ht]
\caption{\textbf{Updating Agent}}
\label{tab:UpdatingAgent}
\begin{response}
\small 
You are now an organization fine-tuner for the software development process. 

Your task is to update the coding agent teams to ensure that the software requirements can be achieved.

The overall requirements, current coding agent teams, and the issues in current implementation of the software are listed below: 

Task and user requirements: "\{task\}".

Requirements: "\{requirements\}"

Coding team composition: "\{composition\}".

Coding team workflow: : "\{workflow\}".

Source Codes: "\{codes\}"

Execution Results: "{test\_reports}" (Note: These results only indicate whether the code has runtime errors and do not reflect functionality correctness.)

Current Issues:
 "\{issues\}".

As the {assistant\_role}, you should give out a detailed plan to update the coding agent teams to ensure that the software requirements can be achieved.

You should first think step by step, reasoning about Requirements, Source Codes, Execution Results, and Current Issues to decide whether the source code has fully accomplished the coding requirements. If here has any conflict, please refer to the Task and user requirements in checking the source code.

Then according to the requirement assessment, you should update the coding agent teams to ensure that the software requirements can be achieved. You can take following actions:

1) Modify the existing coding agent prompts to focus on fixing runtime errors first, then work on completing and perfecting the tasks. Each task should clearly define the issues the agent is expected to solve.

2) Delete coding agents from the WORKFLOW whose requirements have been achieved.

3) Add new coding agent teams in the composition and workflow to ensure that the unimplemented requirements can be achieved.

The requirement assessment should follow the "REQUIREMENTS PROGRESS" format: "achieved" represents whether the requirement has been achieved, and "double-checked" represents whether the requirement has been double-checked. The detailed progress should be included in the answer.

While the updated coding agent team should be in the following format: "COMPOSITION" is the composition of programmers' tasks, and "Workflow" is the workflow of the programmers. Each task is assigned to a programmer, and the workflow shows the dependencies between programmers' tasks.

\#\#\# REQUIREMENTS PROGRESS 

requirement: description of the requirement

achieved: True/False

double-checked: True/False

detailed progress: your analysis

\#\#\# COMPOSITION

```

Programmer 1: Task 1 description - includes specific issues to resolve, necessary code modifications, and expected improvements

Programmer 2: Task 2 description - includes specific issues to resolve, necessary code modifications, and expected improvements

```

\#\#\# WORKFLOW

```

Programmer 1: []

Programmer 2: [Programmer 1]

```

Please note that the coding team should be both effective and efficient.

1) Prioritize fixing runtime errors before addressing requirement issues.

2) Remove the agent from the origin Coding team workflow if the task has been fully completed.

3) The overall tasks should fully cover all uncomplished requirement and current issues. Make sure the Task description as detail as possible. And NEVER include task that modifiy the code to satistify the testing.

4) The composition are not limited to 2 agents. But if there are more than 10 tasks, consider merging the tasks and focus on the most essential features.

5) The workflow should not contain circles!

\end{response}
\end{table}

\section{Software Presentation}
\label{app:software}
In this section, we show some games and websites written by EvoMAC. Fig.~\ref{fig:game gui} and Fig.~\ref{fig:web gui} present the games and websites respectively. We see that: i) EvoMAC outputs games with well-written GUI and game rules. It can handle different kinds of GUI and game rule requirements from diverse games. ii) EvoMAC outputs websites with beautified, user-friendly web pages and correct transition logic. It can handle the requirements of different websites.

\begin{figure}[!t]
    \centering
    \begin{subfigure}[t]{0.55\textwidth}
        \centering
        \includegraphics[height=2.4in]{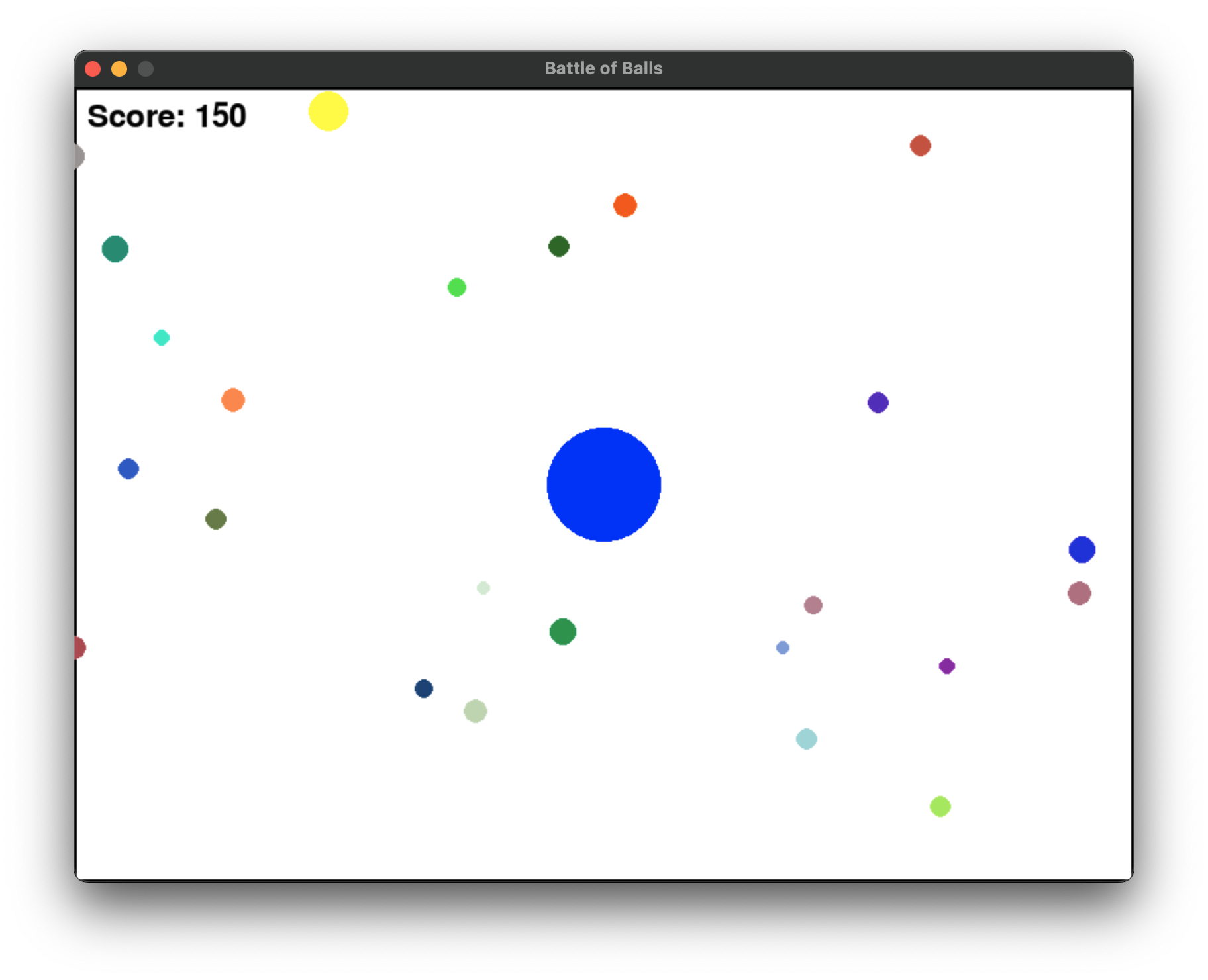}
        \caption{Balls}
    \end{subfigure}
    \begin{subfigure}[t]{0.42\textwidth}
        \centering
        \includegraphics[height=2.4in]{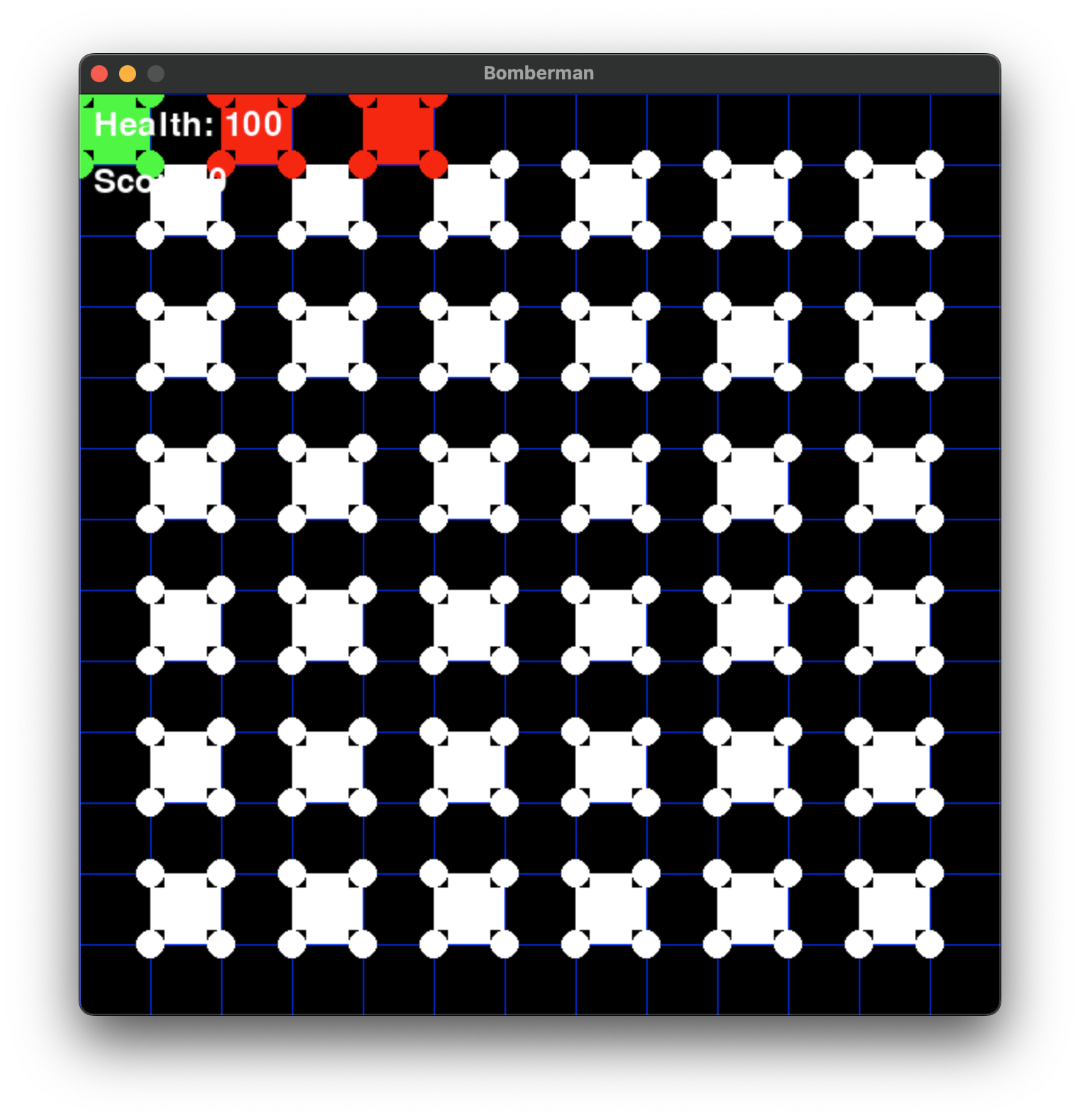}
        \caption{Bomberman}
    \end{subfigure}
    \\
    \begin{subfigure}[t]{0.55\textwidth}
        \centering
        \includegraphics[height=2.4in]{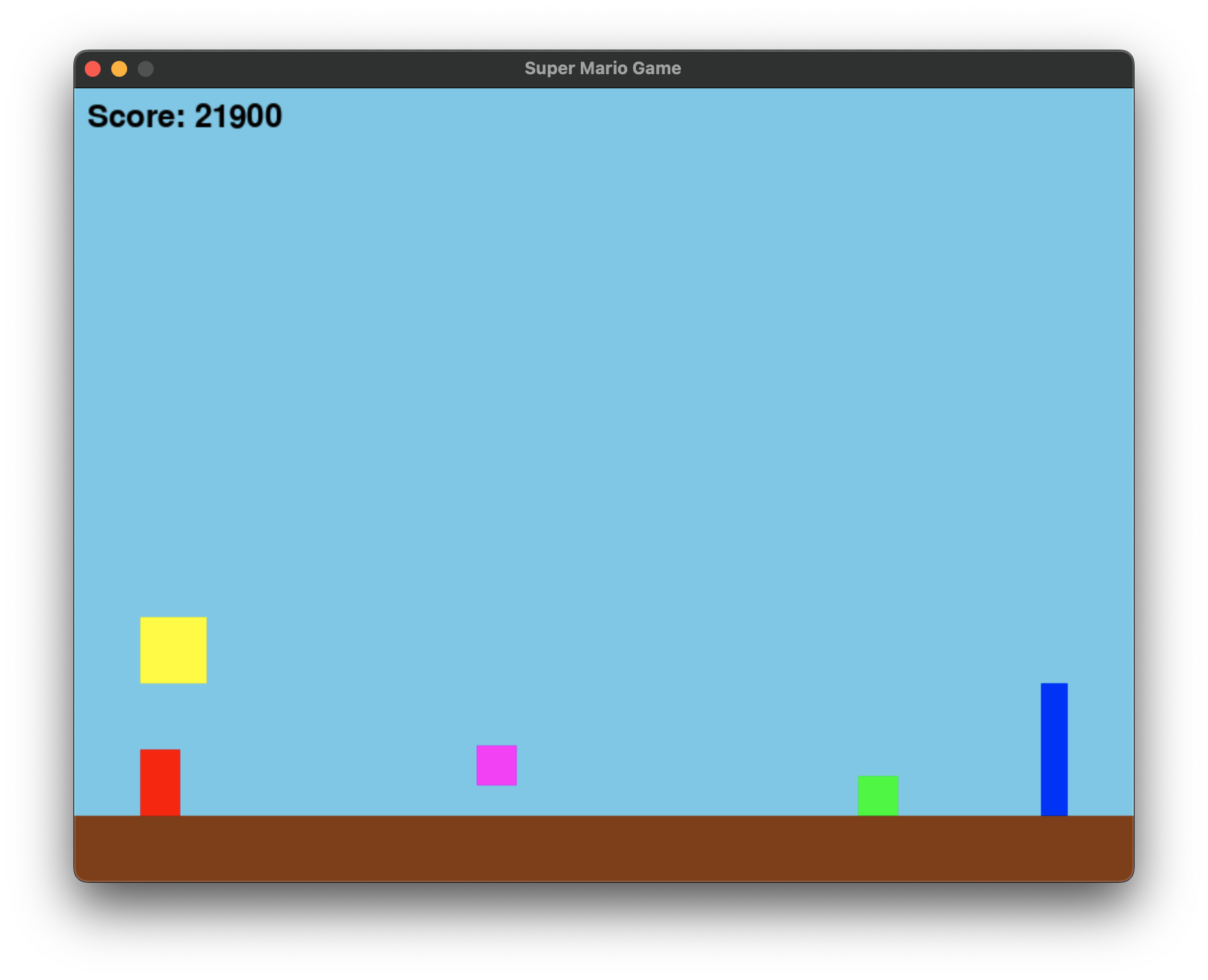}
        \caption{Mario}
    \end{subfigure}%
        \begin{subfigure}[t]{0.42\textwidth}
        \centering
        \includegraphics[height=2.4in]{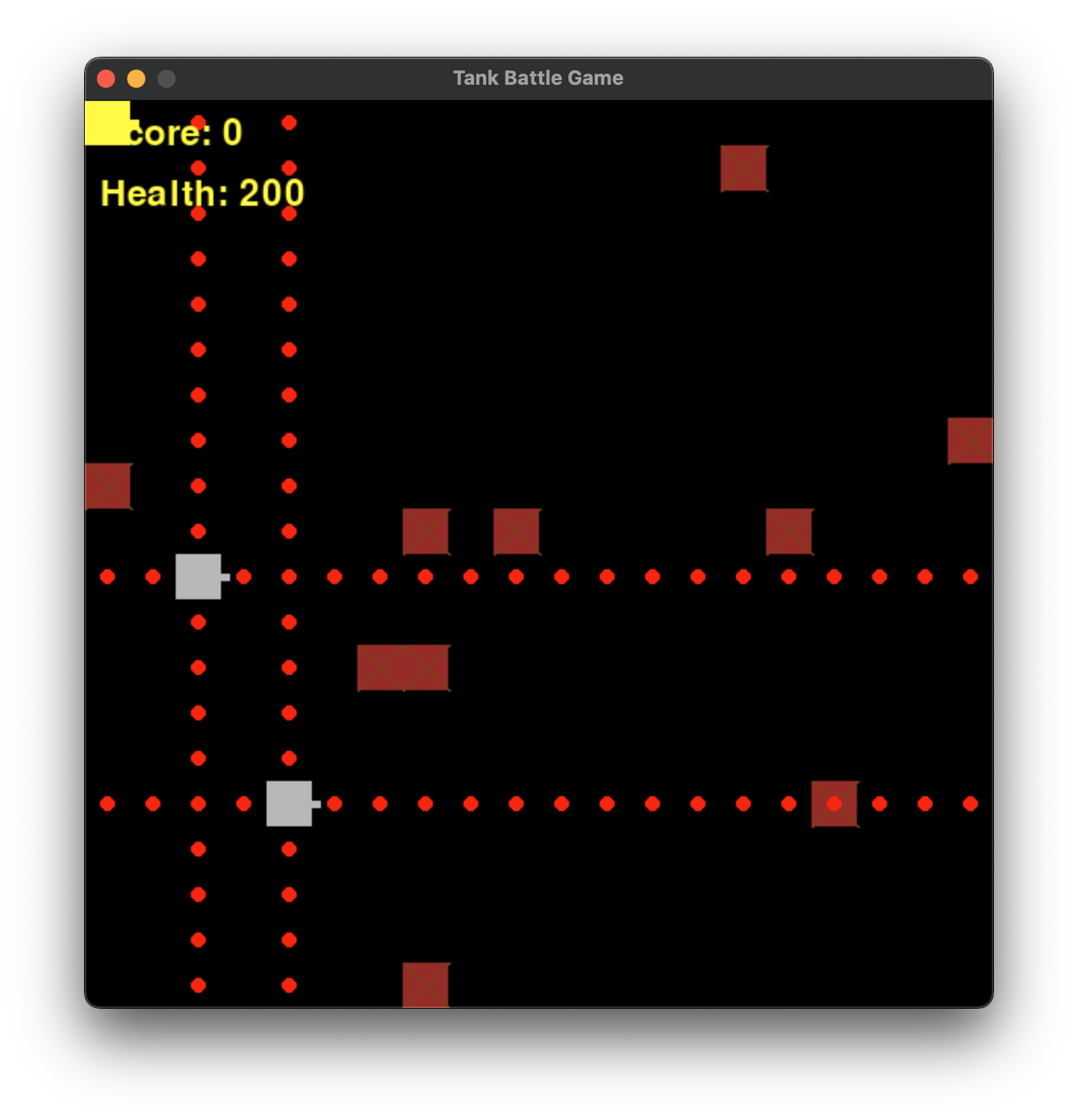}
        \caption{Tank}
    \end{subfigure}%
    \\
    \begin{subfigure}[t]{0.55\textwidth}
        \centering
        \includegraphics[height=2.4in]{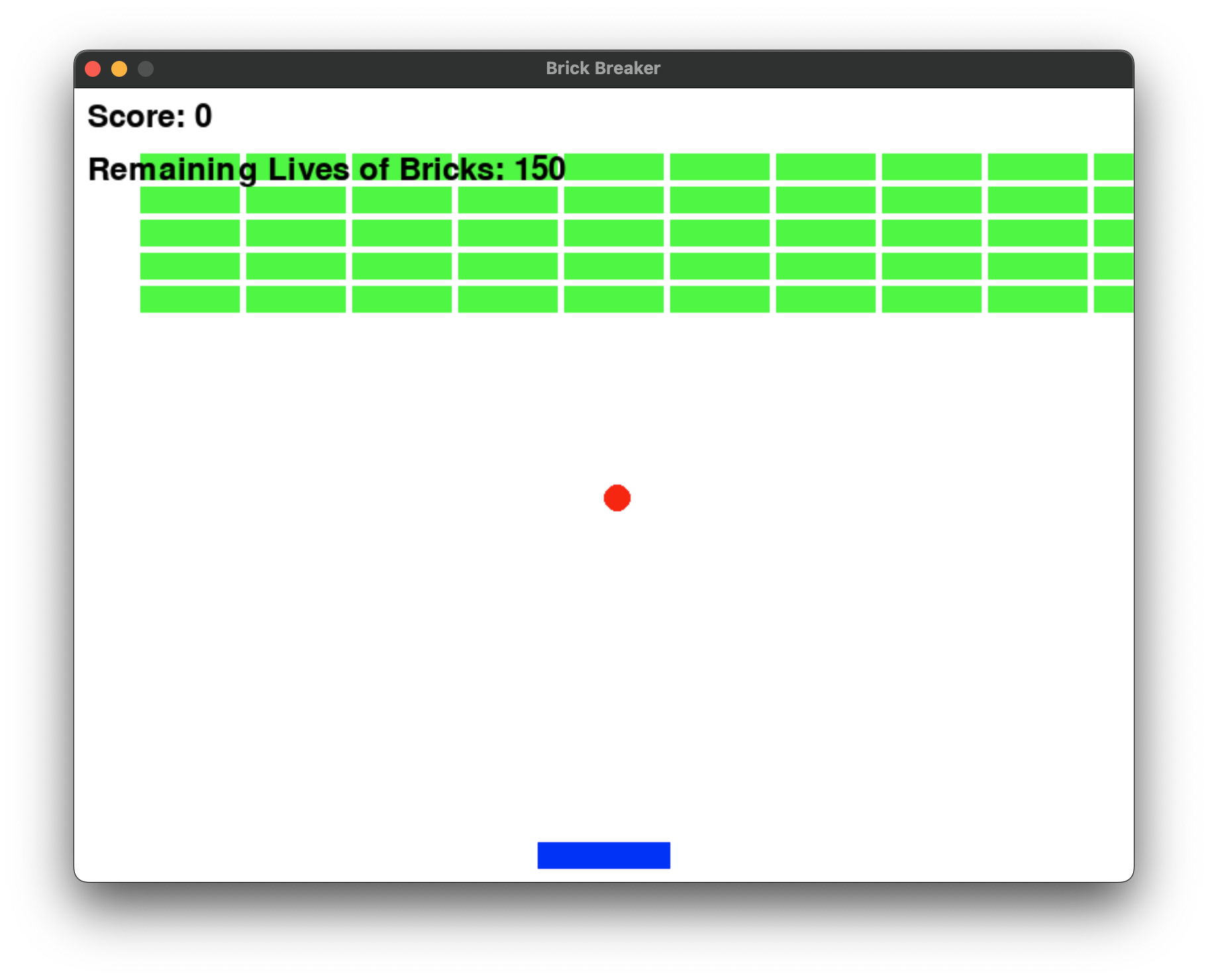}
        \caption{Brick}
    \end{subfigure}%
    \begin{subfigure}[t]{0.42\textwidth}
        \centering
        \includegraphics[height=2.4in]{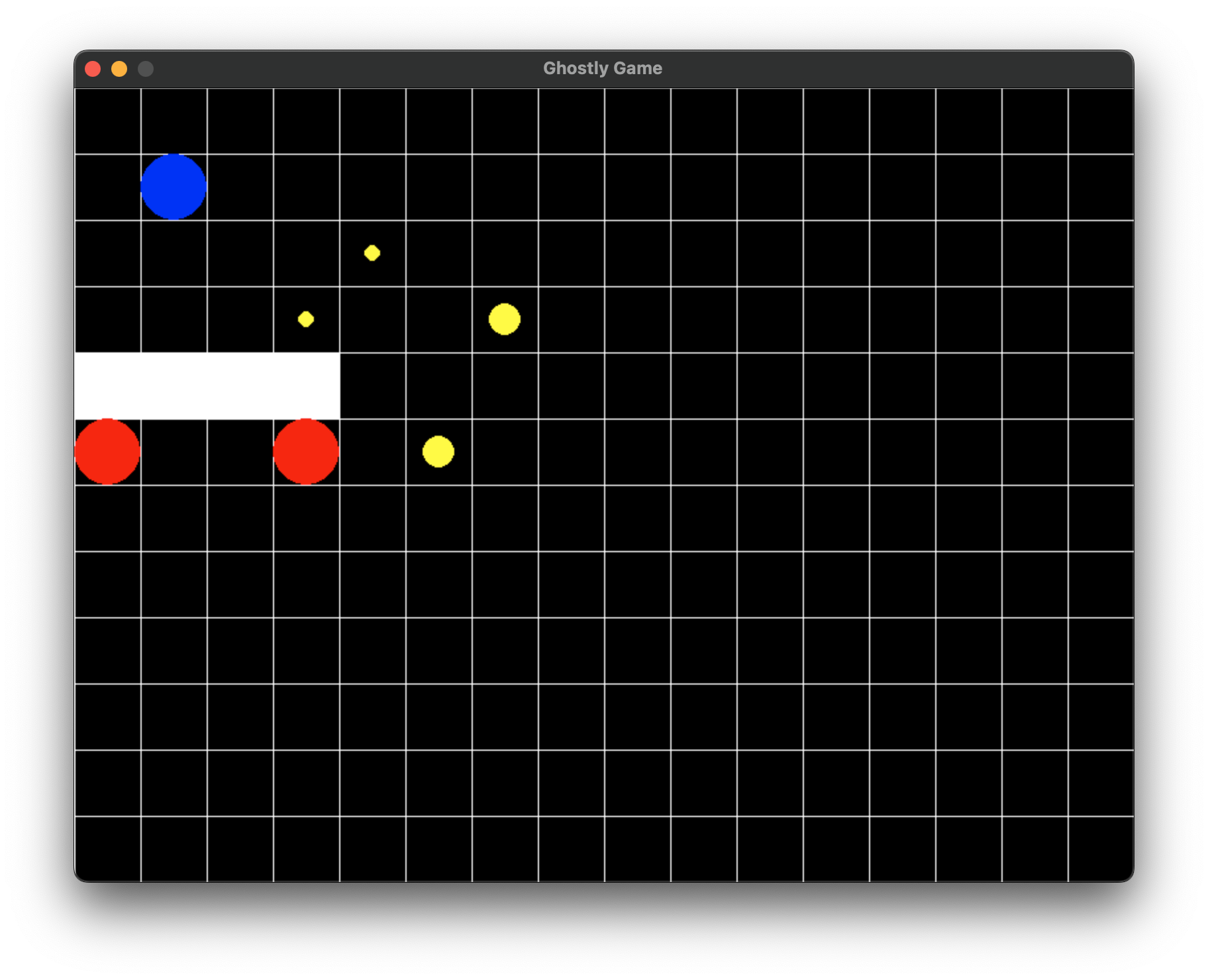}
        \caption{Ghostly}
    \end{subfigure}%
\end{figure}
\begin{figure}[!t]
    \ContinuedFloat
        \begin{subfigure}[t]{0.55\textwidth}
        \centering
        \includegraphics[height=2.4in]{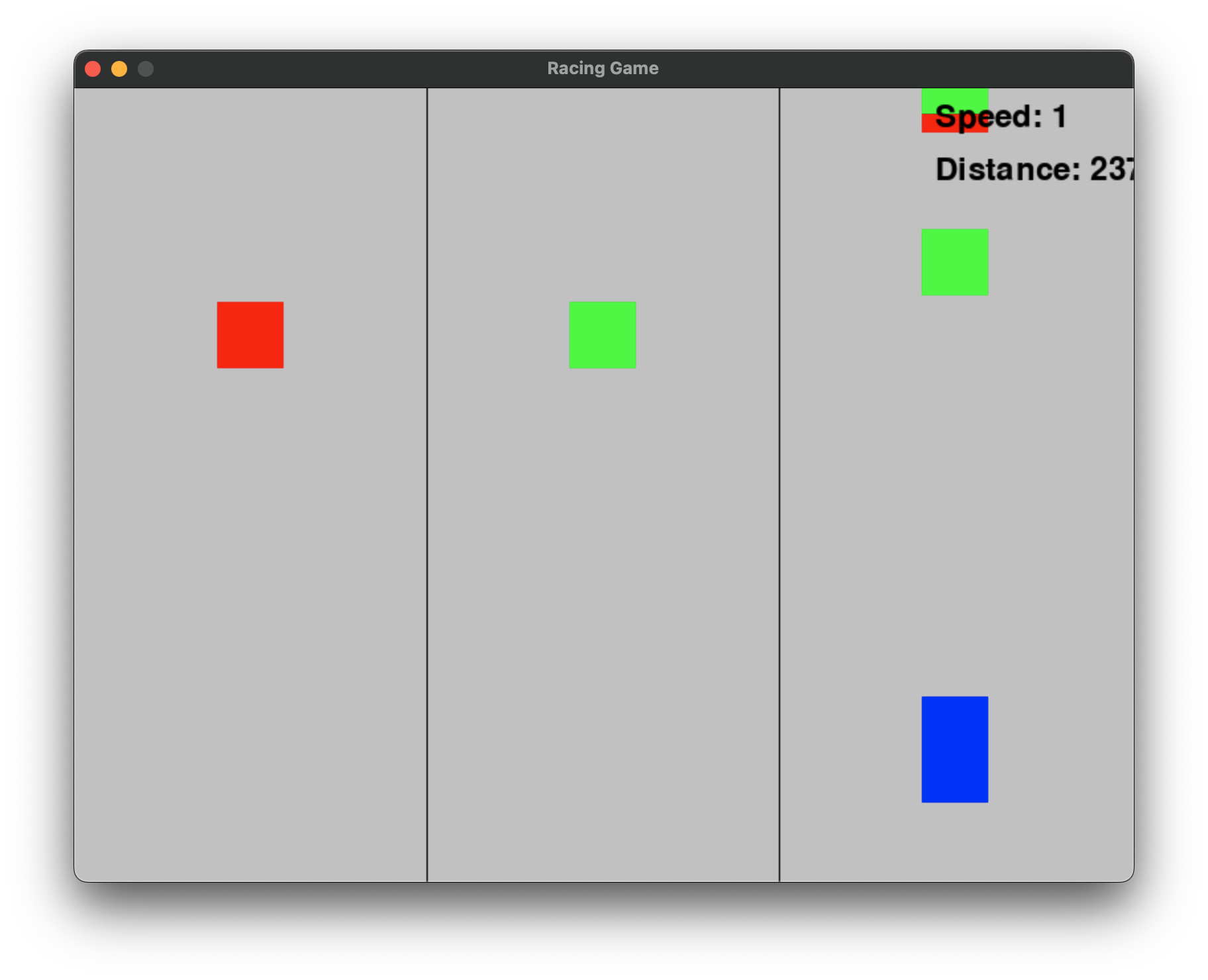}
        \caption{Racing}
    \end{subfigure}%
        \begin{subfigure}[t]{0.42\textwidth}
        \centering
        \includegraphics[height=2.4in]{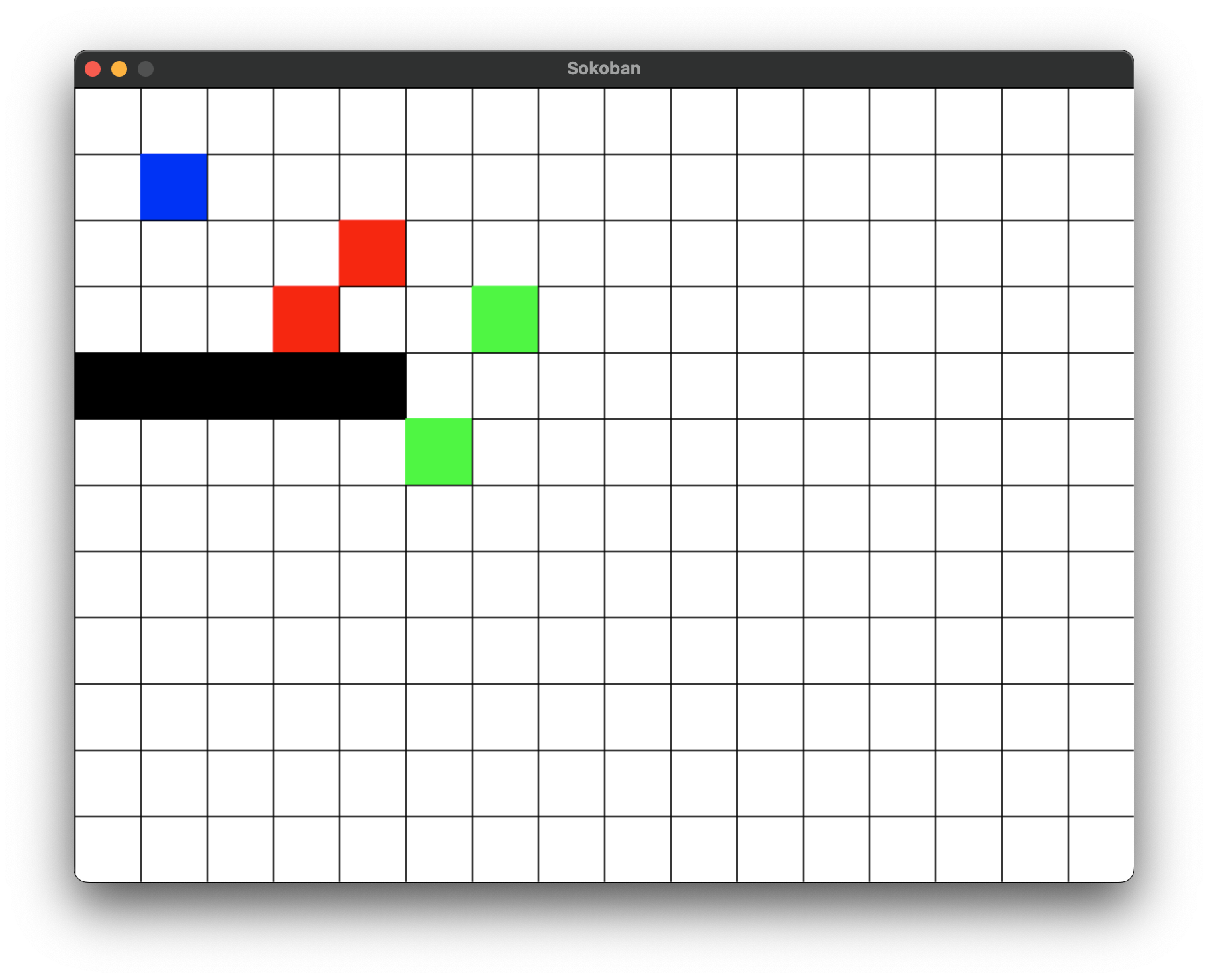}
        \caption{Sokoban}
    \end{subfigure}%
    \caption{Games generated by EvoMAC. } 
    \label{fig:game gui}
\end{figure}


\begin{figure}[!t]
    \centering
    \begin{subfigure}[t]{0.55\textwidth}
        \centering
        \includegraphics[height=2.4in]{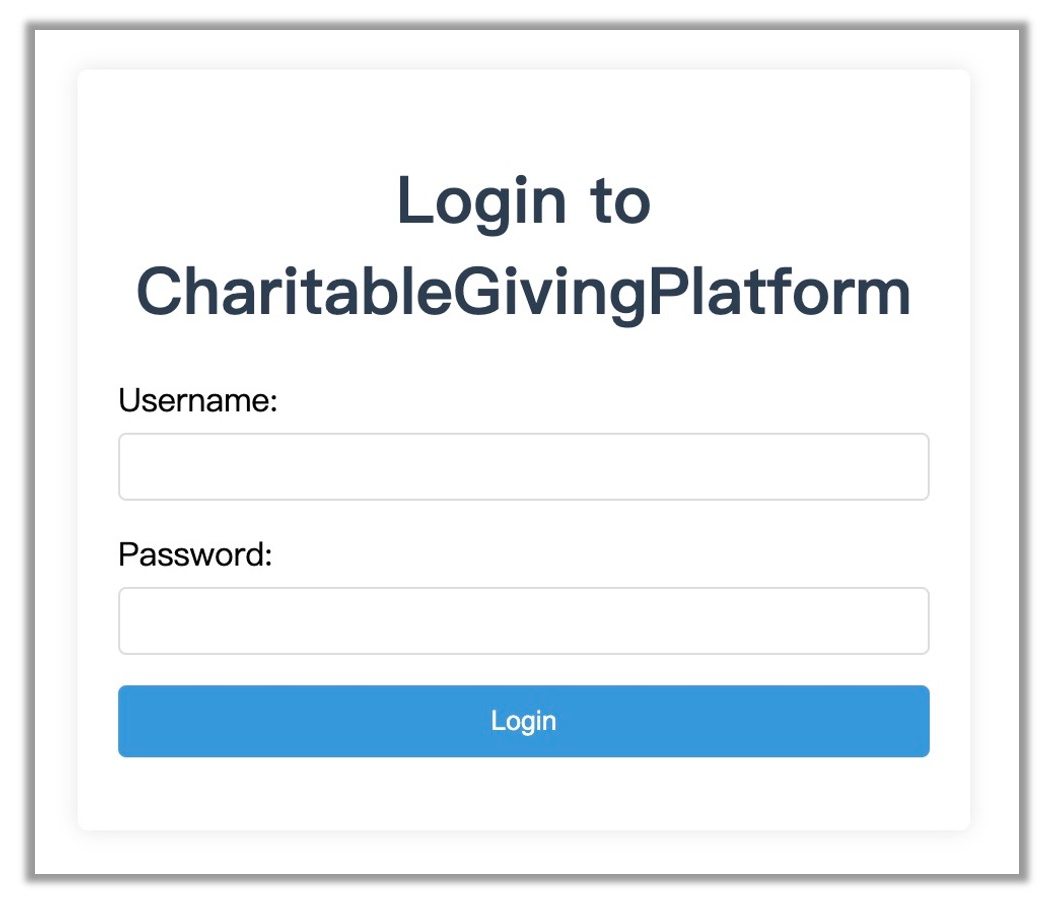}
        \caption{CharitableGivingPlatform}
    \end{subfigure}
    \begin{subfigure}[t]{0.42\textwidth}
        \centering
        \includegraphics[height=2.4in]{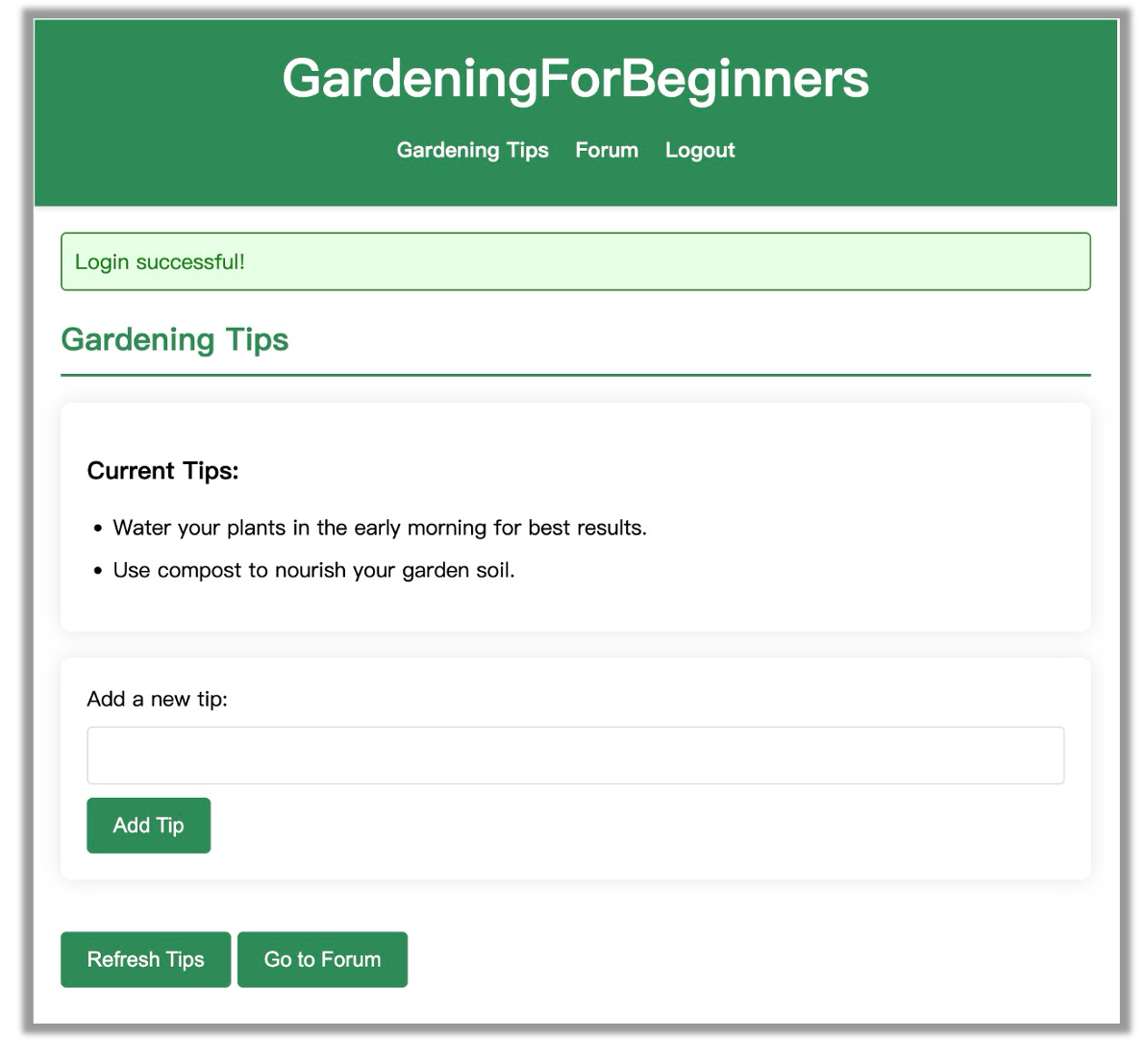}
        \caption{GardeningForBeginners}
    \end{subfigure}%
    \\
    \begin{subfigure}[t]{\textwidth}
        \centering
        \includegraphics[height=2.4in]{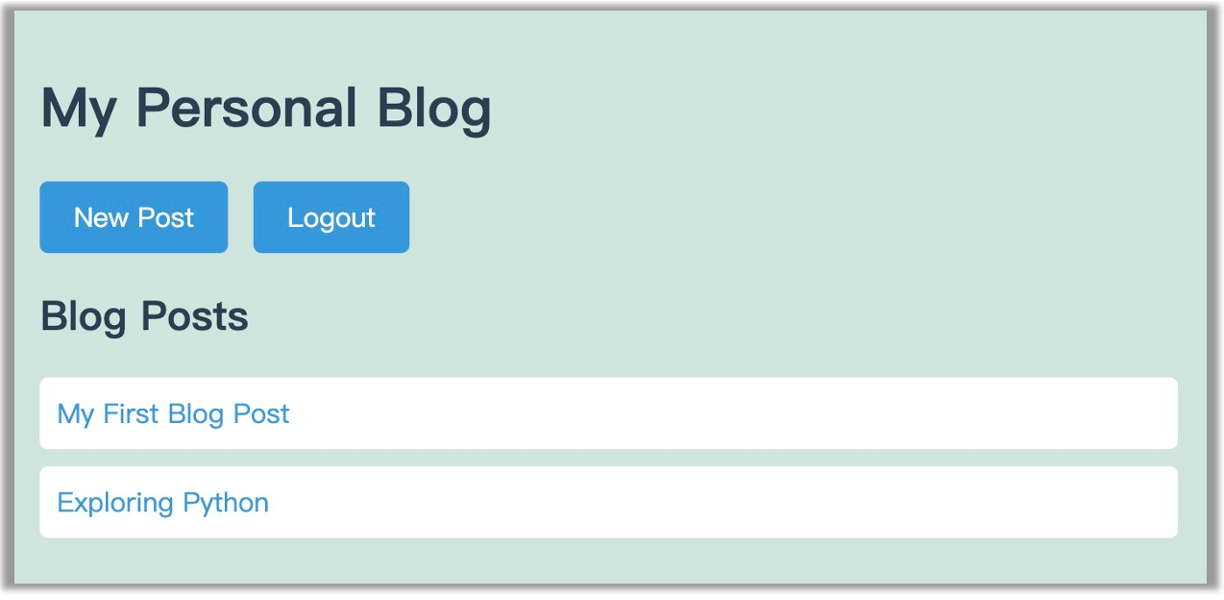}
        \caption{PersonalBlog}
    \end{subfigure}%
\end{figure}
\begin{figure}[!t]
    \ContinuedFloat
    \raggedright
    \begin{subfigure}[t]{0.53\textwidth}
        \centering
        \includegraphics[height=2.4in]{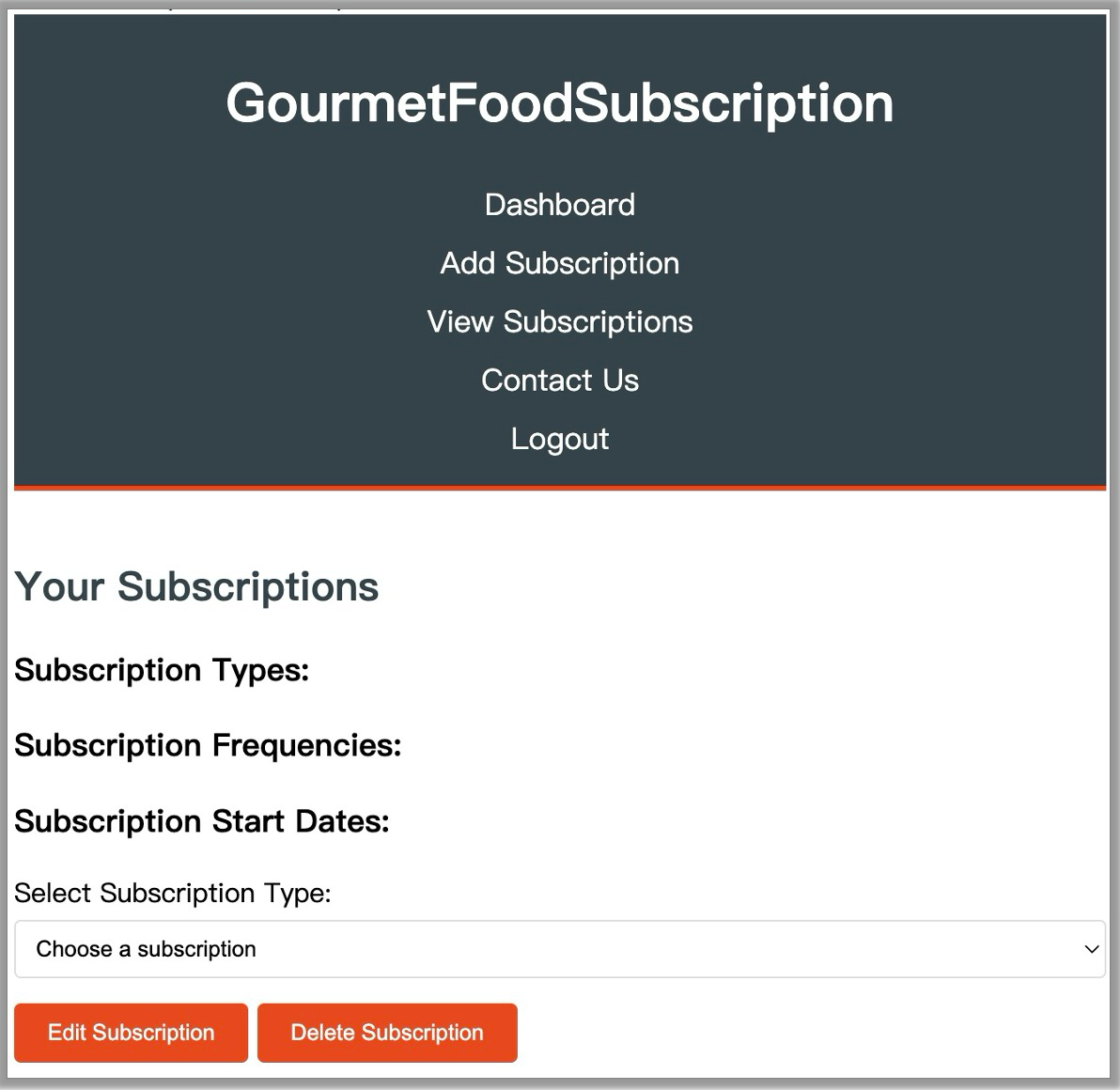}
        \caption{GourmetFoodSubscription}
    \end{subfigure}
    \begin{subfigure}[t]{0.44\textwidth}
        \centering
        \includegraphics[height=2.4in]{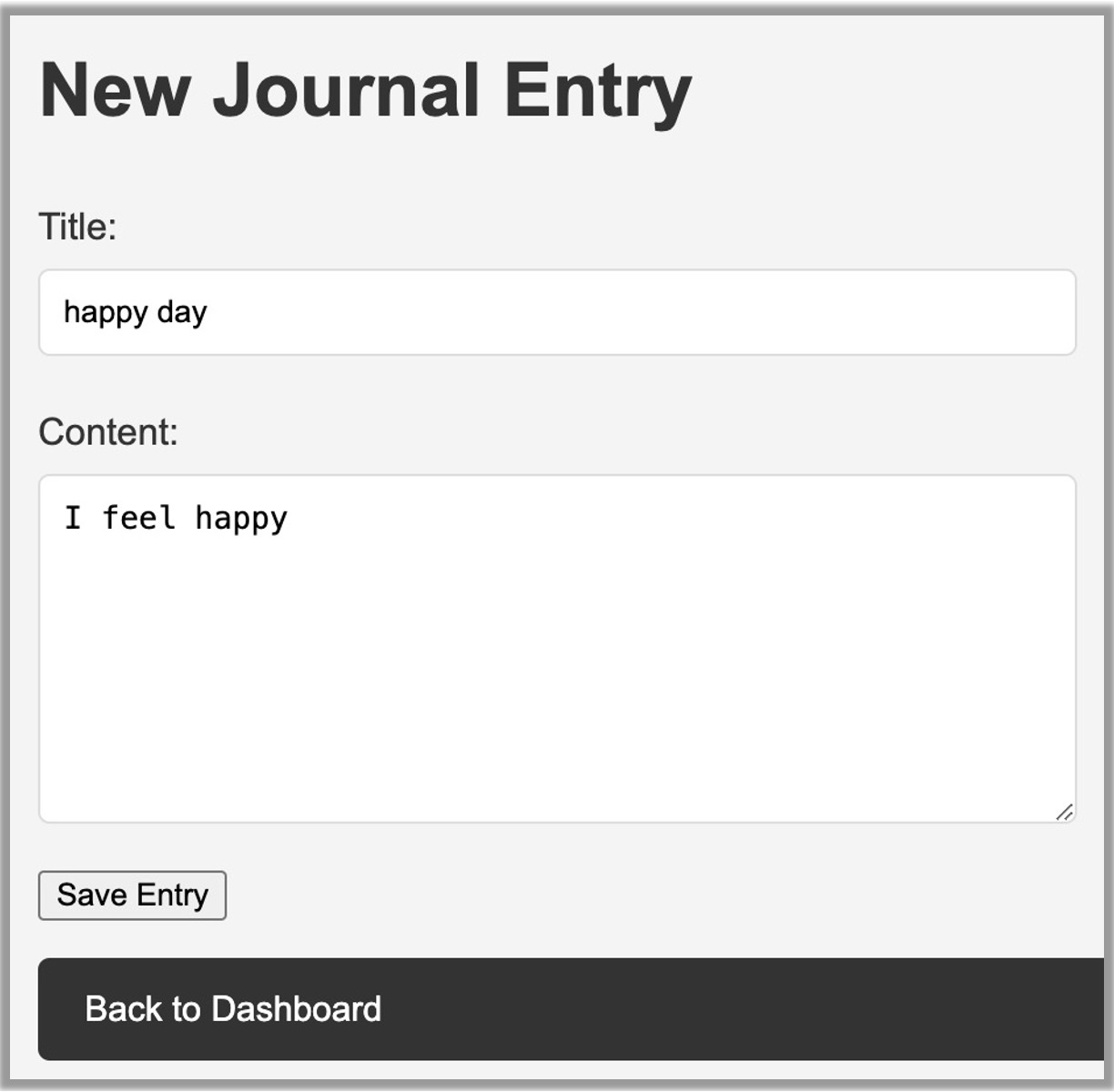}
        \caption{OnlineTherapeuticJournaling}
    \end{subfigure}%
    \\
    \begin{subfigure}[t]{\textwidth}
        \centering  
        \includegraphics[height=2.4in]{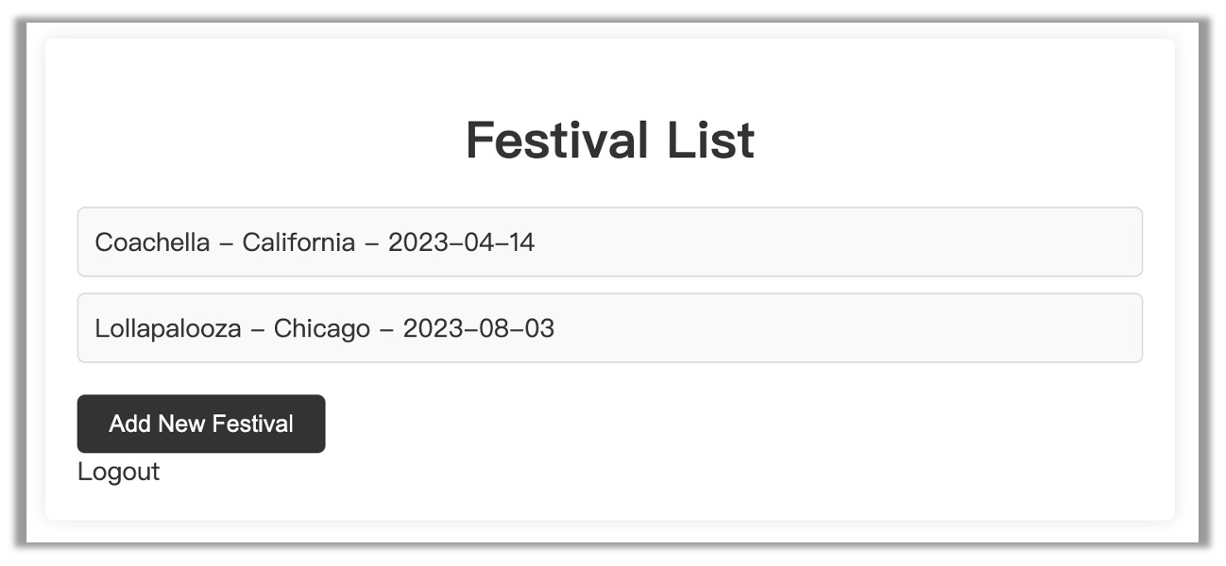}
        \caption{MusicFestivalDirectory}
    \end{subfigure}%
    \caption{Websites generated by EvoMAC. } 
    \label{fig:web gui}
\end{figure}

\end{document}